\documentclass[a4,12pt]{article}

\usepackage[top=1in, bottom=1in, left = 1in, right = 1in]{geometry}
\usepackage{caption, lscape, setspace, soul, tocloft}

\usepackage{amsmath, amssymb, amsthm, mathtools, mathptmx, upgreek}
\usepackage{array, booktabs, multirow} 
\usepackage{makecell}

\usepackage{graphicx} 
\usepackage{adjustbox, float, rotating} 

\usepackage{hyperref}
\usepackage{cleveref}

\usepackage{natbib} % for version 1
\begin{document}
\title{An improved nonparametric test and sample size procedures for the randomized complete block designs}

\author{
  Show-Li Jan\\
  Department of Applied Mathematics\\
  Chung Yuan Christian University\\
  200 Zhongbei Road\\
  Taoyuan 320314, Taiwan\\
  Email: sljan@cycu.edu.tw
\and 
  Gwowen Shieh\thanks{Corresponding author}\\
  Department of Management Science\\
  National Yang Ming Chiao Tung University\\
  1001 University Road\\
  Hsinchu 300093, Taiwan\\
  Email: gwshieh@nycu.edu.tw\\
  \href{https://orcid.org/0000-0001-8611-4495}{ORCID: 0000-0001-8611-4495}
} 

\date{}
\maketitle

\singlespacing
\maketitle
\begin{abstract}
The Friedman test has been extensively applied as a nonparametric alternative to the conventional $F$ procedure for comparing treatment effects in randomized complete block designs. A chi-square distribution provides a convenient approximation to determining the critical values for the Friedman procedure in hypothesis testing. However, the chi-square approximation is generally conservative and the accuracy declines with increasing number of treatments. This paper describes an alternative transformation of the Friedman statistic along with an approximate $F$ distribution that has the same numerator degrees of freedom as the ANOVA $F$ test. Moreover, two approximate noncentral $F$ distributions are presented for the proposed $F$-transformation under the alternative hypothesis of heterogeneous location shifts. Explicit power functions are derived when the underlying populations have the uniform, normal, Laplace, and exponential distributions. Theoretical examination and empirical assessment are presented to validate the advantages of the proposed approaches over the existing methods of the Friedman test. The developed test and power procedures are recommended due to their consistently acceptable Type I error rates and accurate power calculations for the location shift structures and population distributions considered here.
\end{abstract}

\newpage 
\doublespacing
%1
\section{Introduction}
The randomized complete block design (RCBD) and associated ANOVA $F$ procedure are useful for comparing treatment effects by dividing heterogeneous experimental units into homogeneous blocks. Moreover, the distribution-free Friedman test (\citeyear{Friedman1937}) provides a nonparametric alternative to the conventional $F$ procedure if the underlying normal assumption is not tenable. The asymptotic chi-square distribution of the Friedman statistic provides a useful approximation to determining the critical values in hypothesis testing. Despite its simplicity, the conservative nature of the chi-square approximation has been the target of criticism. Exact critical values of the Friedman test are tabulated in \cite{Hollander1999}, \cite{LopezVazquez2019}, Martin, Leblanc and Toan (\citeyear{Martin1993}), \cite{Odeh1977}, \cite{Owen1962}. Computer algorithms are also available to carry out exact calculations or Monte Carlo simulation in Hollander, Wolfe, and Chicken (\citeyear{Hollander2013}) and Schneider, Chicken, and Becvarik (\citeyear{Schneider2016}).

Various rank transform approaches and extensions have been considered as alternatives to the Friedman test for RCBD. \cite{Iman1980} applied the ANOVA $F$ procedure to the intrablock ranks of the observations instead of the observations themselves. The resulting $F$-transformation is assumed to have the same $F$ distribution as the standard ANOVA $F$ test. Accordingly, the hybrid method provides a more accurate approximation than the Friedman test with a chi-square approximation. Also, \cite{Brownie1994} noted that the \cite{Kendall1939} procedure is equivalent to comparing the prescribed $F$-transformation with intrablock ranks to an $F$ distribution with two smaller degrees of freedom than those of the usual ANOVA $F$ distribution. Thus, reduction of numerator and denominator degrees of freedom can help improve the liberal $F$ approximation of \cite{Iman1980} to some extent. Unfortunately, the approximate $F$ distribution of \cite{Kendall1939} turns out to have two fractional degrees of freedom and seems to receive little attention in the literature. No numerical comparison has been presented to illustrate the relative performance of the two $F$ approximations.

Power analyses and sample size calculations are essential for researchers to credibly detect meaningful hypotheses and validate research goals. It is noteworthy that the Friedman test statistic is no longer distribution-free when there are heterogeneous location shifts. To simplify the complex nature, \cite{Hettmansperger1984} presented a noncentral chi-square approximation under the local alternative assumption with small location shifts. Similar assumption was adopted in \cite{Lehmann1998} and \cite{Noether1987} to derive power functions for the Wilcoxon signed-rank test and the Wilcoxon-Mann-Whitney rank sum test. However, the simplified procedures of \cite{Lehmann1998} and \cite{Noether1987} can be quite inaccurate in some cases as shown in Shieh, Jan, and Randles (\citeyear{Shieh2006}, \citeyear{Shieh2007}). The application of such approximations should be restricted to shift models with relatively small departures from the null hypothesis. In the case of RCBD, noncentral chi-square approximation of \cite{Hettmansperger1984} has not been properly examined. Further investigations are required to fully explicate its adequacy for power calculations.

In addition to the intrablock ranks, there are rank transform, aligned ranks and other ranking schemes in which ranks can be assigned to observations in RCBD. The related discussions and reviews were demonstrated in \cite{Brownie1994}, \cite{Conover1981}, Iman, Hora, and Conover (\citeyear{Iman1984}), \cite{Lehmann1998}, and \cite{Mahfoud2005}, among others. Owing to the wide acceptance of Friedman’s (\citeyear{Friedman1937}) test, this study is concerned with the intrinsic issues and useful extensions of Friedman statistic based on intrablock ranks. This paper attempts to improve the current rank procedures for RCBD in two respects. First, following the principles of \cite{Box1955}, \cite{Pitman1938}, and \cite{Welch1937}, a general class of $F$-transformations and approximate $F$ distributions is described and accordingly, it encompasses the \cite{Kendall1939} procedure as a special case. More importantly, a prominent $F$-transformation is also identified and the associated approximate $F$ distribution has the same numerator degrees of freedom as the ANOVA $F$ test. Second, under the alternative hypothesis with heterogeneous location shifts, noncentral $F$ distributions are suggested for the $F$-transformations to facilitate power evaluation and sample size determination. For practical assessments, explicit formulations are derived when the underlying populations have the uniform, normal, Laplace, and exponential distributions. Extensive numerical results reveal that the recommended test procedure and approximate nonnull $F$ distribution have obvious advantages over the Friedman test with a noncentral chi-square approximation for improving the control of Type I error rates and accuracy of power computations.

%2
\section{Test procedures}
The usual model for randomized complete block designs is defined as follows:
\begin{equation} %(1)
    X_{ij} = \upmu+\uptheta_i+\upgamma_j+\upvarepsilon_{ij},
\end{equation}
where $X_{ij}$ is the response of the $i$th treatment in the $j$th block, $\upmu$ is the overall mean, $\uptheta_i$ is the $i$th treatment effect, $\upgamma_j$ is the $j$th block effect, and the errors $\upvarepsilon_{ij}$ are independent and identically distributed for $i = 1, \, \ldots, \, K \ (\geq 2)$ and $j = 1, \, \ldots, \, B \ (\geq 2)$. In addition, the effects of $\uptheta_i$ and $\upgamma_j$ are constants subject to the restriction $\sum\limits_{i=1}^K \uptheta_i = \sum\limits_{j=1}^B \upgamma_j = 0$. Accordingly, the responses $X_{ij}$ are assumed to have the continuous distribution functions $F_{ij}(x)=F(x-\upmu-\uptheta_i-\upgamma_j),$ and the question of interest is whether there are differences among the additive treatment effects in terms of the hypotheses
\begin{equation} %(2)
    H_0: \mathrm{All} \ \uptheta_i \ \mathrm{are \ zero \ versus} \ H_1: \mathrm{At \ least \ one} \ \uptheta_i \ \mathrm{differs \ from \ zero}.
\end{equation}
The null hypothesis asserts that the underlying distributions $F_{ij}$ within block $j$ are the same $F_{1j} = F_{2j} = \ldots = F_{Kj}$ for each fixed $j = 1, \, \ldots, B$.

%2.1
\subsection{\textit{Two popular procedures}}
\cite{Friedman1937} proposed a distribution-free test based on within-block rankings $R_{ij}$ that represents the rank assigned to $X_{ij}$ within block $j$. The Friedman statistic $T$ is then given by
\begin{equation} %(3)
    T = \frac{12}{BK(K+1)} \sum\limits_{i=1}^K R_i^2 - 3B(K+1),
\end{equation}
where $R_i=\sum\limits_{j=1}^B R_{ij}$ denotes the sum of ranks of the $i$th treatment. When the ranks within each group have the same integer or $R_{i1} = R_{i2} = … = R_{iB}$ for $i = 1, \, \ldots , \, K,$ it follows that $\sum\limits_{i=1}^K R_i^2 = B^2 \sum\limits_{i=1}^K i^2 = B^2K(K + 1)(2K + 1)/6$ and $T$ attains the maximum $M = B(K - 1)$. Under the null hypothesis, it was shown in \cite{Friedman1937} that the statistic $T$ has an asymptotic chi-square distribution with $K - 1$ degrees of freedom $T \ \dot{\sim} \ \upchi_{K - 1}^2$. Accordingly, the null hypothesis is rejected if $T > \upchi_{K - 1, \, \upalpha}^2$ where $\upchi_{K - 1, \, \upalpha}^2$ is the upper $(100\cdot\upalpha)$th percentile of the $\upchi_{K - 1}^2$ distribution. The chi-square approximation of \cite{Friedman1937} is usually conservative and deteriorates as the number of treatments $K$ increases.

Alternatively, \cite{Conover1981} suggested a direct extension of the ANOVA $F$ statistic for testing equal treatment effects with the intra-block ranks $R_{ij}$. In this setting, the block sum of squares $SSBL = 0$. The total sum of squares, the treatment sum of squares and the error sum of squares have the simple form $SSTO = BK(K^2 - 1)/12,$ $SSTR = K(K + 1)T/12$ and $SSE = K(K + 1)(M - T)/12,$ respectively. Consequently, the resulting statistic $F_R$ is a monotone transformation of the Friedman’s statistic
\begin{equation} %(4)
    F_R = \frac{(B-1)T}{M-T}.
\end{equation}
The rank-based statistic $F_R$ is assumed to have the same $F$ distribution as the parametric counterpart $F_R \ \dot{\sim} \  F(r_1, \, r_2),$ where $r_1 = K - 1,$ $r_2 = (B - 1)(K - 1),$ and $F(r_1, \, r_2)$ is an $F$ distribution with degrees of freedom $r_1$ and $r_2$. Hence, this rank test rejects the null hypothesis of no differences among treatment effects at the significance level $\upalpha$ if $F_R > F_\upalpha(r_1, \, r_2)$ where $F_\upalpha(r_1, \, r_2)$ is the upper $(100\cdot\upalpha)$th percentile of the $F(r_1, \, r_2)$ distribution. The exact and simulation comparisons in \cite{Iman1980} showed that the critical value $F_\upalpha(r_1, \, r_2)$ is usually smaller than the exact one so that the $F$ approximation of $F_R$ is liberal. But the degree of oversize is less than the conservative level of the chi-square approximation of $T$ according to the marginal $10\%$ deviation of nominal Type I error rate. An $F$ distribution $F(a, b)$ is asymptotically identical to $\upchi_{a}^2/a$, and $F(a, b)$ provides a reasonable approximation of $\upchi_{a}^2/a$ for finite sample sizes. Therefore, it is theoretically and empirically supported that $F$-type transformation deserves critical recognition and further investigations. Following the moment-matching principles of Box and Andersen (1955), Pitman (1938), and Welch (1937), improved procedures are described next to give better Type I error control than the liberal $F_R$ test. 

%2.2
\subsection{\textit{A general class of rank procedures}}
To provide alternative rank procedures, the techniques of \cite{Box1955}, \cite{Pitman1938}, and \cite{Welch1937} are extended as a key approach to constructing feasible tests. Accordingly, the suggested test procedures are formulated in two steps. First, the essential idea is to construct a general class of rank transformed statistics and approximated distributions. Specifically, the desirable formulation is obtained by matching the mean and variance of the scaled statistic $B_S$ with those of a Beta distribution so that $B_S \ \dot{\sim} \ Beta(f_1/2, \, f_2/2)$ where $B_S = T/S$ and $S$ is a crucial factor to be specified in the next step. Note that, under the null hypothesis, the mean and variance of the $T$ statistic are $\upmu_{0T} = K - 1$ and variance $\upsigma_{0T}^2 = 2(K - 1)(B - 1)/B,$ respectively. The derivations imply that $f_1 = \mu_{0T}d,$ $f_2 = (S - \upmu_{0T})d,$ and $d = (2/S)\{\upmu_{0T}(S - \upmu_{0T})/\upsigma_{0T}^2 - 1\}$. Using the relationship between a Beta distribution and an $F$ distribution, it is convenient to consider the monotonic function $F_S$ of $B_S$:
\begin{equation} %(5)
    F_S = \frac{T/f_1}{(S-T)/f_2} \ \dot{\sim} \ F(f_1, \, f_2).
\end{equation}
Notably, the proposed formulation $F_S$ and associated approximate $F$ distribution give a unified class of useful test procedures in terms of $(S, \, f_1, \, f_2)$ where $S$ is a function of $K,$ $B,$ $\upmu_{0T}$, and $\upsigma_{0T}^2$.

Subsequently, two special cases are identified from the class of approximate $F$ procedures to provide desirable tests for detecting treatment effects. The first procedure is to replace $S$ with $M$ for the statistic $F_S$ and the distribution $F(f_1, \, f_2)$.

It is straightforward to obtain the special case
\begin{equation} %(6)
    F_M = \frac{(B-1)T}{M-T} \ \dot{\sim} \ F(m_1, \, m_2),
\end{equation}
where $m_1 = K - 1 - 2/B$ and $m_2 = (B - 1)m_1$. At the significance level $\upalpha,$ this procedure rejects the null hypothesis if $F_M > F_\upalpha (m_1, \, m_2)$. As noted in \cite{Brownie1994}, the rank procedure $F_M$ along with the approximate distribution $F(m_1, \, m_2)$ has been described in \cite{Kendall1939} from a different perspective. Note that the statistic $F_M$ is the same as the prescribed $F_R$. However, the degrees of freedom of the two approximate $F$ distributions differ. The small difference between $m_1$ and $r_1$ suggests that $F_\upalpha (m_1, \, m_2) > F_\upalpha (r_1, \, r_2)$ and the $F_M$ test generally has a smaller rejection rate than the $F_R$ test. But the examination in \cite{Iman1980} and related comparisons did not cover the $F_M$ method. It is of great interest to demonstrate the actual discrepancy between the adjusted degrees of freedom and usual degrees of freedom.

Note that the degrees of freedoms $m_1$ and $m_2$ are fractional. The common tables of an $F$ distribution only present the quantiles of integer degrees of freedom. The utility of $F_M$ is apparently limited by the task of finding the $F$ quantiles with two fractional degrees of freedom. Conceivably, it is temping to consider a transformation $F_S$ with both degrees of freedom $f_1$ and $f_2$ are integer values. A direct and intuitive choice of $f_1$ is the salient number $K – 1$ as the degrees of freedom for $T \ \dot{\sim} \ \upchi_{K - 1}^2$. This consideration provides an important motivation to identify and examine the following viable member $F_L$ of the general class $F_S$.

%2.3
\subsection{\textit{The proposed procedure}}
An alternative approach is to bring the approximation $F_S \ \dot{\sim} \ F(f_1, \, f_2)$ into closer relation to the usual ANOVA $F$ by fixing the numerator degrees of freedom $f_1$ as $l_1 = K - 1$ or, equivalently, $d = 1$. Then, the quantity $S$ can be derived as $L = B(K + 1) - 2$ and the denominator degrees of freedom $f_2$ has the specific form $l_2 = (B - 1)(K + 1)$. The suggested test statistic and approximate distribution are formulated as
\begin{equation} %(7)
    F_L = \frac{(K+1)(B-1)T}{(K-1)(L-T)} \ \dot{\sim} \ F(l_1, \, l_2).
\end{equation}
Accordingly, the test rejects the null hypothesis of no treatment effects if $F_L > F_\upalpha (l_1, \, l_2)$ at the significance level $\upalpha$. Interestingly, the approximations $F_L$ and $F_R$ have the same numerator degrees of freedom $l_1 = r_1 = K - 1,$ but the other matching quantities are different as $L > M$ and $l_2 > r_2$. Moreover, unlike the fractional degrees of freedoms $m_1$ and $m_2$ of $F_M$, both $l_1$ and $l_2$ of $F_L$ are integer values. The Type I error control of the suggested test and existing procedures will be investigated in the numerical study.

%2.4
\subsection{\textit{Comparison of Type I error rates}}
The approximate null distributions of the rank procedures provide a simple alternative to find the critical points other than the exact values. To validate these approximate procedures, exact assessment and simulation study were conducted to clarify the underlying properties of the $F_R,$ $F_M$ and $F_L$ test procedures in Type I error control.

First, the approximate $F$ distributions of procedures $F_R,$ $F_M$ and $F_L$ are compared with the chi-square approximation of $T$ through the tabulated exact critical values of $T$ in \cite{Hollander1999}. Using the exact critical value $c_\upalpha$ of $T,$ with $P\{T > c_\upalpha\} = \upalpha$ closes to 0.10, 0.05 and 0.01, the error rates are computed for the four approximations:

$P_{Tc} = P\{ \upchi_{K-1}^2 > c_\upalpha \}$;

$P_{Rc} = P\{ F(r_1, \, r_2) > F_{Rc} \}$ with $F_{Rc} = \{ (B-1)c_\upalpha \} / (M-c_\upalpha)$;

$P_{Mc} = P\{ F(m_1, \, m_2) > F_{Mc} \}$ with $F_{Mc} = \{ (B-1)c_\upalpha \} / (M-c_\upalpha)$; and

$P_{Lc} = P\{ F(l_1, \, l_2) > F_{Lc} \}$ with $F_{Lc} = \{ (K+1)(B-1)c_\upalpha \} / \{ (K-1)(L-c_\upalpha) \}$.
\\
Accordingly, the accuracy of an approximation is evaluated by the magnitude of error rate deviates from the designated probability of Type I error. The differences between the computed error rates $(P_{Tc}, \, P_{Rc}, \, P_{Mc}, \, P_{Lc})$ and exact $\upalpha$ are presented for $K = 3$, 4, and 5 in Tables 1-3, respectively. The number of blocks is $B = 3, \, \ldots ,$ 8, and a total of 53 combined structures are considered. For ease of comparison, the percentage of error is also computed as $100(P - \upalpha)/\upalpha$ for $P = P_{Tc},$ $P_{Rc},$ $P_{Mc},$ and $P_{Lc}$. The results in Table 1 for $K = 3$ show that the four contending tests do not have a clear tendency and consist behavior. However, it can be seen from Tables 2 and 3, the conservative chi-square approximation of $T$ generally gives positive errors because of $c_\upalpha < \upchi_{K - 1, \, \upalpha}^2,$ whereas the other three $F$ approximations tend to be liberal and yield negative errors for $K = 4$ and 5. The absolute values of error and percentage of error suggest that all three $F$ procedures outperform the traditional chi-square approximation for Friedman’s $T$. More importantly, although the $F$ approximations are reasonably good, the suggested procedure $F_L$ compares favorably with the $F_M$ and $F_R$ methods.

Second, the test procedures are also examined for several additional configurations when the exact calculation of critical value is computationally difficult. Simulation study were conducted to appraise the Type I error rates for the number of block are large: $B = 5$, 10, 15, and 20. Simulated Type I error probabilities $(P_T, \, P_R, \, P_M, \, P_L)$ of the four rank procedures were computed with 100,000 iterations of standard normal variables for $\upalpha = 0.10$, 0.05 and 0.01, where $P_T = P\{T > \upchi_{K – 1, \, \upalpha}^2 \},$ $P_R = P\{F_R > F_\upalpha (r_1, \, r_2)\},$ $P_M = P\{F_M > F_\upalpha (m_1, \, m_2)\},$ and $P_L = P\{F_L > F_\upalpha (l_1, \, l_2)\}$. Accordingly, the errors and percentages of errors between the simulated probability and the nominal value of $\upalpha$ for a total of 36 cases are presented in Tables 4, 5, and 6 for $K = 3$, 4 and 5, respectively. In these cases, the chi-square approximation of $T$ commonly incurs negative errors because of $c_\upalpha < \upchi_{K – 1, \, \upalpha}^2$  or $P_T < \upalpha$. However, the liberal $F_R$ procedure often has positive errors and such phenomena are especially evident for the 24 cases of $K = 4$ and 5. Similar to the previous results, the performance of the chi-square test is slightly inferior to the other three procedures. The empirical results reveal that the new approach $F_L$ possesses some advantages over the $F_R$ and $F_M$ procedures in maintaining good Type I error rate.

%2.5
\subsection{\textit{An example}}
To illustrate the intrinsic differences of alternative procedures for detecting treatment effects, the breaking strength study in \citet[p. 108]{Cochran1957} is reexamined here. The experiment was conducted to examine the effects of potash level in the soil on the breaking strength of cotton fibers. Specifically, five levels of potash ($K = 5$) were applied in a randomized block design with three blocks ($B = 3$). The response was the Pressley strength index, which measures the breaking strength of a bundle of fibers of a designated section. According to the means of four determinations made from the sample of cotton from each plot, the rank sums of the five specified Potash levels are $R_1 = 5, \, R_2 = 5, \, R_3 = 9, \, R_4 = 14,$ and $R_5 = 12$. It can be readily shown that the Friedman statistic $T = 8.8$ with degrees of freedom $K - 1 = 4$ and $p$-value = 0.0663. The prescribed three $F$ statistics are $F_R = 5.5, \, F_M = 5.5,$ and $F_L = 3.6667$ with degrees of freedom $(r_1, \, r_2) = (4, \, 8), \, (m_1, \, m_2) = (3.33, \, 6.67),$ and $(l_1, \, l_2) = (4, \, 12),$ and the associated $p$-values are 0.0199, 0.0301, and 0.0357, respectively. The results show that the traditional Friedman test fails to reject the null hypothesis of no treatment effects at the significance level 0.05. Whereas all the other three transformed $F$ test procedures reject the null hypothesis and demonstrate that there is certain treatment effects among the five levels of potash.

%3
\section{Power calculations} 
%3.1
\subsection{\textit{Fundamental results under the null hypothesis}}
The fundamental properties of the Friedman’s rank statistic are presented here for the sake of completeness and convenient reference. First, under the null hypothesis, it can be shown that the mean $E[R_{ij}] = (K + 1)/2,$ variance $Var[R_{ij}] = (K^2 - 1)/12,$ and covariance $Cov(R_{ij}, \, R_{lj}) = -(K + 1)/12$ for $i \neq l$. These results immediately lead to that the mean $E[R_i] = \upmu_{0i},$ variance $Var[R_i] = \upsigma_{0i}^2,$ and covariance $Cov(R_i, \, R_l) =  \upsigma_{0il}$ of the rank sum of the $i$th treatment $R_i = \sum\limits_{j = 1}^B R_{ij}$ where $\upmu_{0i} = B(K + 1)/2, \, \upsigma_{0i}^2 = B(K^2 - 1)/12,$ and $\upsigma_{0il} = -B(K + 1)/12$ for $i \neq l$. Moreover, standard derivations show the critical results that $\upmu_{0T} = E[T] = K - 1$ and $\upsigma_{0T}^2 = Var[T] = 2(K - 1)(B - 1)/B$ as presented in \cite{Friedman1937} and \cite{Hettmansperger1984}.

To obtain the asymptotic distribution of $T,$ a useful and alternative expression is $T = \sum\limits_{i = 1}^K T_i^2$ where $T_i = (1 - 1/K)^{1/2}\{R_i\, –\, \upmu_{0i}\}/ \upsigma_{0i}$. Note that $E[T_i] = u_{0i} = 0, \, Var[T_i] = v_{0i}^2 = v_{0ii} = 1 - 1/K,$ and $Cov(T_i, \, T_l) = v_{0il} = -1/K$ for $i \neq l$. Using the large-sample theorem, it can be shown that
\begin{equation} %(8)
    \mathbf{T} = (T_1, \, \ldots, \, T_K)^\mathrm{T} \ \dot{\sim} \ N_K(\mathbf{u}_0, \, \mathbf{V}_0),
\end{equation}
where $\mathbf{u}_0 = (u_{01}, \, \ldots, \, u_{0K})^\mathrm{T} = \mathbf{0}_K$ is a $K \times 1$ null vector and $\mathbf{V}_0 = \{v_{0il}\}$ is the $K \times K$ variance-covariance matrix of $\mathbf{T}$. Note that $\mathbf{V}_0$ is idempotent with rank $(K - 1)$. It follows from the theorem for the quadratic forms of the asymptotic multivariate normal vectors in \citet[Section 3.5]{Serfling1980} that $T = \mathbf{T}^\mathrm{T}\mathbf{T}$ has the asymptotic distribution $\upchi_{K - 1}^2$ under the null hypothesis. With slightly different arguments, the particular property was also noted in \citet[Exercise 4.5.7.]{Hettmansperger1984}, and interestingly the proof presented in \cite{Friedman1937} was adapting from the original result by Dr. S. S. Wilks.

%3.2
\subsection{\textit{Location shift formulation}} 
Under the location shift assumption that at least one $\uptheta_i$ differs from the others for $F_{ij}(x)=F(x-\upmu-\uptheta_i-\upgamma_j),$ the properties of the rank sum statistics and Friedman statistic are more complex than those under the null hypothesis. Accordingly, a detailed discussion of the nonnull issues does not exist to our knowledge. For ease of exposition, explicit expressions are derived and presented here. It follows that the expected value $\upmu_i = E[R_i]$ of the rank sum statistic is $\upmu_i = B\upmu_{ij}$ where $\upmu_{ij} = E[R_{ij}] = \sum\limits_{l \neq i}^K P_{1il} + 1$ and $P_{1il} = P\{X_{i1} > X_{l1}\} = \int F(x+\uptheta_i-\uptheta_l)\mathrm{d}F(x)$. Let $\mathbf{u} = E[\mathbf{T}] = (u_1, \, \ldots, \, u_K)^T$ where $u_i = (1 - 1/K)^{1/2}\{\upmu_i - \upmu_{0i}\}/ \upsigma_{0i}$ for $i = 1, \, \ldots, \, K$. Accordingly, both the quantity $P_{1il}$ and the effect size $\uptau = \mathbf{u}^\mathrm{T}\mathbf{u}$ depend on the underlying distribution $F_{ij}(x) = F(x - \upmu - \uptheta_i - \upgamma_j)$. The variance of $R_i$ can be shown as $\upsigma_i^2 = B\upsigma_{ij}^2,$ where
\begin{gather*}
    \upsigma_{ij}^2 = Var[R_{ij}] = \sum\limits_{l \neq i}^K P_{1il} - \{ \sum\limits_{l \neq i}^K P_{1il} \}^2 + 2\sum\limits_{l \neq i}^K \ \sum\limits_{l < m \neq i}^K P_{2i(l, \, m)} \ \mathrm{and}
    \\ 
    P_{2i(l, \, m)} = P\{ X_{i1}>X_{l1}, \, X_{i1}>X_{m1} \} = \int F(x+\uptheta_i-\uptheta_l)F(x+\uptheta_i-\uptheta_m)\mathrm{d}F(x).
\end{gather*}
Moreover, the expected value of the Friedman test statistic $\upmu_T = E[T]$ is
\begin{equation} %(9)
    \upmu_T = \frac{12}{BK(K+1)} \sum\limits_{i=1}^K (\upmu_i^2+\upsigma_i^2) - 3B(K+1),
\end{equation}
because of $E[R_i^2] = \upmu_i^2 + \upsigma_i^2$ and $\upsigma_i^2 = Var[R_i]$. Apparently, $\upmu_T$ is a function of the mean and variance $(\upmu_i, \, \upsigma_i^2)$ of the rank sum statistics $R_i$, and in turn, they depend on the underlying distribution $F_{ij}(x)$. Hence, the distribution-free nature of the Friedman test under the null hypothesis does not generalize to the nonnull situation.

Due to technical complexity, \cite{Hettmansperger1984} considered a simple Taylor expansion of $P_{1il}$ at $(\uptheta_i, \, \uptheta_l) = (0, \, 0)$ and it suggests that $P_{1il} \doteq  1/2 + f^*(0)(\uptheta_i - \uptheta_l)$ where $f^*(0) = \int f^2(x)\mathrm{d}x$ and $f(\cdot)$ is the probability density function. Moreover, under the specific formulation of local alternative $\uptheta_i = \upbeta_i/(BK)^{1/2},$ \citet[Exercise 4.5.8]{Hettmansperger1984} showed that
\begin{equation} %(10)
    \upmu_i \doteq B(K+1)/2+(BK)^{1/2}f^*(0)(\upbeta_i - \bar{\upbeta}) \ \mathrm{and} \ u_i=E[T_i] \doteq \{ 12/(K+1) \}^{1/2} f^*(0)(\upbeta_i - \bar{\upbeta}),
\end{equation}
where $\bar{\upbeta}=\sum\limits_{i=1}^K \upbeta_i/K$ (With the notation here, $\sum\limits_{i=1}^K \uptheta_i = \sum\limits_{i=1}^K \upbeta_i = 0$ and $\bar{\upbeta} = 0$). Accordingly, $\uptau$ is readily simplified as
\begin{equation} %(11)
    \uptau_H = \sum\limits_{i=1}^K u_i^2 = \{ 12/(K+1) \} \{ f^*(0) \}^2 \sum\limits_{i=1}^K (\upbeta_i - \bar{\upbeta})^2.
\end{equation}

Furthermore, the variance-covariance matrix $\mathbf{V} = Var[\mathbf{T}]$ also varies with the population distribution $F_{ij}(x)$ and does not generally have a simplified form and idempotent feature as the counterpart $\mathbf{V}_0$ under the null hypothesis. To obtain a tractable approximation, \cite{Hettmansperger1984} assumed under the local alternative assumption with small location shifts that $\mathbf{V} \doteq \mathbf{V}_0$ and $\mathbf{T} \ \dot{\sim} \ N_K(\mathbf{u}, \, \mathbf{V}_0)$. Consequently, \cite{Hettmansperger1984} suggested that $T$ has the nonnull asymptotic distribution $T \ \dot{\sim} \ \upchi_{K-1}^2 (\uptau_H)$ where $\upchi_{K-1}^2 (\uptau_H)$ is a noncentral chi-square distribution with degrees of freedom $(K - 1)$ and noncentrality $\uptau_H$. The associated power function of the Friedman test is
\begin{equation} %(12)
    \Psi_H = P\{ \upchi_{K-1}^2 (\uptau_H) > \upchi_{K-1, \, \upalpha}^2 \}.
\end{equation}

In contrast to the derivation and approximation presented by \cite{Hettmansperger1984}, an intuitive approach is to extend the null chi-square distribution of $T$ to a noncentral chi-square $\upchi_{K-1}^2 (\uptau_A)$ under the alternative hypothesis where the noncentrality parameter $\uptau_A$ is yet to be properly identified. Specifically, a natural consideration is to match the first moments of the distribution $\upchi_{K-1}^2 (\uptau_A)$ and the test statistic $T$. It is clear from $E[\upchi_{K-1}^2 (\uptau_A)] = K - 1 + \uptau_A$ and $\upmu_T = E[T]$ given in Equation 9 that $\uptau_A \doteq \upmu_T - (K - 1)$. It follows from the local alternative simplification of \cite{Hettmansperger1984} that $\upmu_i \doteq B(K +1)/2 + (BK)^{1/2}f^*(0)(\upbeta_i - \bar{\upbeta}), \, \upsigma_i^2 \doteq \upsigma_{0i}^2 = B(K^2 - 1)/12,$ and $\upmu_T \doteq K - 1 + \uptau_H$. This implies that the noncentrality parameter $\uptau_A$ of the presumed chi-square distribution is $\uptau_A \doteq \uptau_H$ and this leads to the same asymptotic approximation $T \ \dot{\sim} \ \upchi_{K - 1}^2 (\uptau_H)$ as shown above.

%3.3
\subsection{\textit{The proposed nonnull distributions and power functions}}
The current nonnull approximation proposed in \cite{Hettmansperger1984} has an attractive and simplified expression. However, the particular formulation was valid only under some special structure and condition. More importantly, the corresponding properties in power calculations do not appear to have been previously investigated in the literature. Further empirical examinations revealed that the noncentral chi-square approximation does not provide a generally reliable power function of Friedman’s test under the location shift scenario. More accurate approaches and distributions should be considered.

Naturally, a noncentral $F$ distribution is considered for the general distribution of $F_S$: $F_S \ \dot{\sim} \ F(f_1,$ $f_2, \, \updelta_S)$ where $F(f_1, \, f_2, \, \updelta_S)$ is a noncentral $F$ distribution with degrees of freedom $(f_1, \, f_2)$ and noncentrality $\updelta_S$. Using $E[T] = \upmu_T, \, E[f_1F_S] = E[f_2T/(S - T)] \doteq f_2\updelta_S^*$ where $\updelta_S^* = \upmu_T/(S - \upmu_T)$. Standard asymptotic theorems assure that $\upchi_\upnu^2/\upnu$ approaches 1 as $\upnu$ goes to infinity and $f_1F(f_1, \, f_2, \, \updelta_S) = \upchi_{f_1}^2(\updelta_S)/(\upchi_{f_2}^2/f_2) \doteq \upchi_{f_1}^2(\updelta_S)$ for large sample sizes. Thus, $E[f_1F(f_1, \, f_2, \, \updelta_S)] \doteq E[\upchi_{f_1}^2(\updelta_S)] = \updelta_S + f_1$. By equating the two quantities of $E[f_1F_S]$ and $E[f_1F(f_1, \, f_2, \, \updelta_S)],$ a sensible noncentrality parameter for $\updelta_S$ is $\updelta_{SA} = f_2\updelta_S^* - f_1$. Note that $T$ and $(S - T)$ are negatively correlated and the quantity $\updelta_S^*$ underestimates the mean value $E[T/(S - T)]$ to some extent. Therefore, an inflating factor can be applied to obtain a more accurate approximation. In contrast to the asymptotic approximation $\upchi_\upnu^2/\upnu \doteq 1,$ direct derivations show that $E[\upnu/\upchi_\upnu^2] = \upnu/(\upnu\, –\, 2)$ for $\upnu > 2$. Accordingly, the adjusting factor $f_2/(f_2\, -\, 2)$ is applied to $\updelta_S^*$ or $E[f_1F_S] \doteq \{f_2^2/(f_2 – 2)\} \updelta_S^*$ under finite sample consideration. A revised approximation for $\updelta_S$ is $\updelta_{SB} = \{f_2^2/(f_2 - 2)\} \updelta_S^* - f_1$.

The prescribed identification shows that $F_M$ and $F_L$ are two notable members of the $F_S$ class. With the suggested noncentral $F$ approximation $F(f_1, \, f_2, \, \updelta_S)$ for $F_S,$ the rank transform $F_M$ has the noncentral $F$ approximations:
\begin{equation} %(13)
    F_M \ \dot{\sim} \ F(m_1, \, m_2, \, \updelta_{MA}) \ \mathrm{and} \ F_M \ \dot{\sim} \ F(m_1, \, m_2, \, \updelta_{MB}),
\end{equation}
where $\updelta_{MA} = m_2\updelta_M^* - m_1, \, \updelta_{MB} = \{m_2^2/(m_2 - 2)\}\updelta_M^* - m_1,$ and $\updelta_M^* = \upmu_T/(M - \upmu_T)$. For ease of illustration, the corresponding power functions are denoted by
\begin{equation} %(14)
    \Psi_{MA} = P\{ F(m_1, \, m_2, \, \updelta_{MA}) > F_\upalpha(m_1, \, m_2) \} \ \mathrm{and} \ 
    \Psi_{MB} = P\{ F(m_1, \, m_2, \, \updelta_{MB}) > F_\upalpha(m_1, \, m_2) \}, 
\end{equation}
respectively. Moreover, the $F_L$ statistic is assumed to have the noncentral $F$ distributions
\begin{equation} %(15)
    F_L \ \dot{\sim} \ F(l_1, \, l_2, \, \updelta_{LA}) \ \mathrm{and} \ F_L \ \dot{\sim} \ F(l_1, \, l_2, \, \updelta_{LB}), 
\end{equation}
where $\updelta_{LA} = l_2\updelta_L^* - l_1, \, \updelta_{LB} = \{l_2^2/(l_2 - 2)\}\updelta_L^* - l_1,$ and $\updelta_L^* = \upmu_T/(L - \upmu_T)$. The associated power functions are expressed as
\begin{equation} %(16)
    \Psi_{LA} = P\{ F(l_1, \, l_2, \, \updelta_{LA}) > F_\upalpha(l_1, \, l_2) \} \ \mathrm{and} \ 
    \Psi_{LB} = P\{ F(l_1, \, l_2, \, \updelta_{LB}) > F_\upalpha(l_1, \, l_2) \}, 
\end{equation}
respectively.

It is essential to note that the noncentral distribution $\upchi_{K-1}^2 (\uptau_H)$ and power function $\Psi_H$ of the Friedman test presented in \cite{Hettmansperger1984} depend on the population distribution $F_{ij}(x)$ only through $f^*(0) = \int f^2(x)\mathrm{d}x$. It can be shown that the quantity $f^*(0) = 1$, $1/(2\uppi^{1/2})$, 1/4, and 1/2 for uniform, normal, Laplace, and exponential distributions, respectively. In contrast, the underlying properties of $F_{ij}(x)$ are heavily involved in the proposed noncentral $F$ distributions and power functions. Accordingly, the described power formulas $\{\Psi_{MA}, \, \Psi_{MB}, \, \Psi_{LA}, \, \Psi_{LB}\}$ are functions of the probability evaluations: $P_{1il}$ and $P_{2i(l, \, m)}$. The explicit expressions of probability evaluations for the four underlying population distributions of uniform, normal, Laplace, and exponential distributions are derived and presented in the appendix. The performance of the power function $\Psi_H$ for the Friedman test will be examined and compared with the suggested power functions for detecting heterogeneous location shifts in the subsequent numerical study.

%3.4
\subsection{\textit{Power assessments}}
To illustrate the power behavior of the contending formulas, numerical assessments are carried out under a wide range of model configurations. The Type I error rate and nominal power are fixed as $\upalpha = 0.05$ and $1 - \upbeta = 0.90$ throughout this numerical study. The underlying populations are assumed to have four different distributions of uniform, normal, Laplace, and exponential shift alternatives and the number of groups is $K = 3$ and 5. For $K = 3,$ the location shifts are set as $(\uptheta_1, \, \uptheta_2, \, \uptheta_3) = \upsigma\cdot(-1, \, 0, \, 1), \, \upsigma\cdot(-2/3, \, 0, \, 2/3), \, \upsigma\cdot(-1/2, \, 0, \, 1/2),$ and $\upsigma\cdot(-1/3, \, 0, \, 1/3)$. The locations shifts of $K = 5$ are $(\uptheta_1, \, \uptheta_2, \, \uptheta_3, \, \uptheta_4, \, \uptheta_5) = \upsigma\cdot(-1, \, -1/2, \, 0, \, 1/2, \, 1), \, \upsigma\cdot(-2/3, \, -1/3, \, 0, \, 1/3,$ $2/3), \, \upsigma\cdot(-1/2, \, -1/4, \, 0, \, 1/4, \, 1/2),$ and $\upsigma\cdot(-1/3, \, -1/6, \, 0, \, 1/6, \, 1/3)$. Note that the selected uniform, normal, Laplace, and exponential distributions have the variance $\upsigma^2 = 1/12$, 1, 2, and 1, respectively.

The power examinations are conducted in two steps. First, under the designated distribution and location shifts, the minimum numbers of blocks and estimated power levels are computed by the $\Psi_H, \, \Psi_{MA}, \, \Psi_{MB}, \, \Psi_{LA}, \, \Psi_{LB}$ formulas. For ease of illustration, these power procedures based on the formulas $\Psi_H, \, \Psi_{MA}, \, \Psi_{MB}, \, \Psi_{LA}, \, \Psi_{LB}$ are denoted as the $T_H, \, F_{MA}, \, F_{MB}, \, F_{LA}, \, F_{LB}$ approaches, respectively. The results of the four different location shifts with two numbers of groups are summarized in Tables 7-14 for the uniform, normal, Laplace, and exponential distributions, respectively. Second, the simulated power of the Friedman test and the four $F$ transform procedures were computed via Monte Carlo simulation of 100,000 independent data sets for the specified model configurations. For ease of comparison, the simulated power and the difference between the estimated power and simulated power are also presented in Tables 7-14.

As can be seen from these tables, the computed numbers of blocks increase with decreasing location shifts in order to attain the same nominal power 0.90. The optimal number of block also determines the minimum total sample size $N_T = KB$ for the designated structure. Although it is not entirely consistent, the optimal numbers of the five methods have the general order $T_H \leq F_{MB} \leq F_{MA} \leq F_{LB} \leq F_{LA}$. The obvious difference between the noncentrality parameters $\updelta_{MA} \leq \updelta_{MB}$ leads to the sample size relationship that $F_{MB} \leq F_{MA}$. Similarly, the sample sizes between the two procedures $F_{LB} \leq F_{LA}$ can be attributed to $\updelta_{LA} \leq \updelta_{LB}$ for the noncentrality parameters. It is essential to note that the estimated numbers of blocks of the four noncentral $F$ procedures are not substantial and they are within the range of 4 units. Also, the estimated power levels are marginally larger than 0.90 because of the discrete nature of the computed numbers.

Notably, the $T_H$ method requires the least sample size to attain the nominal power, but it tends to incur the largest error among the procedures. The resulting errors are quite pronounced for small sample sizes. Specifically, the largest error in each table is 0.1007, 0.1333, 0.0550, 0.0735, 0.2196, 0.2814, 0.6516, and 0.6916 in Tables 7-14, respectively. Although the differences between the estimated power and simulated power decrease for larger number of blocks, the $T_H$ method does not appear to be a reliable method for general use. On the other hand, the noncentral $F$ approximations are more accurate in power calculations than the noncentral chi-square method. For the two magnitudes of noncentrality, it is interesting to note that $F_{MA}$ is better than $F_{MB}$ whereas $F_{LA}$ is outperformed by $F_{LB}$. Although it does not show complete dominance, $F_{LB}$ has some advantages over the $F_{MA}$, especially when the numbers of blocks are small. For the cases of largest shifts $\upsigma\cdot(–1, \, 0, \, 1)$ and $\upsigma\cdot(–1, \, –1/2, \, 0, \, 1/2, \, 1)$, the resulting error of $F_{MA}$ is 0.0373, 0.0046, 0.0362, 0.0180, 0.0406, 0.0267, 0.0587, and 0.0362 in Tables 7-14, respectively. As presented in the tables, the corresponding error of $F_{LB}$ is –0.0207, –0.0098, –0.0280, –0.0065, –0.0088, 0.0061, –0.0175, and 0.0111, respectively. To visualize the power differences of the examined procedures, they are plotted in Figures 1 and 2 for the largest location shifts $(\uptheta_1, \, \uptheta_2, \, \uptheta_3) = \upsigma\cdot(-1, \, 0, \, 1)$ and $(\uptheta_1, \, \uptheta_2, \, \uptheta_3, \, \uptheta_4, \, \uptheta_5) = \upsigma\cdot(-1, \, -1/2, \, 0, \, 1/2, \, 1)$, respectively. In short, the performance of the $F_{LB}$ approach is reasonably good for small sample sizes. Along with the outstanding accuracy for moderate and large sample sizes, the power investigations suggest that the $F_{LB}$ procedure provides a useful technique for power and sample size calculations.

%3.5
\subsection{\textit{An application}}
The numerical results in the breaking strength study of \cite{Cochran1957} are extended to demonstrate the planning of sample sizes for future designs. The sample treatment effects and sample standard deviation are computed to obtain the standardized treatment effects ($-1.3096$, $-1.0055$, 0.0993, 0.6276, 1.5882). Accordingly, the location shifts are chosen as $\upsigma\cdot(-1.3096$, $-1.0055$, 0.0993, 0.6276, 1.5882) in this numerical demonstration. For the significance level 0.05 and nominal power 0.80, the optimal numbers of blocks of the $F_{LB}$ approach are computed as 5, 4, 4, and 4 for uniform, normal, Laplace, and exponential distributions, respectively. Due to the small sizes, the respective estimated powers can be moderately larger than 0.80 and the actual values are 0.8878, 0.8070, 0.8475, and 0.8737. To achieve the nominal power 0.9, the necessary numbers of blocks are 6, 5, 5, and 5 for the uniform, normal, Laplace, and exponential distributions, respectively. Again, the respective estimated powers are 0.9466, 0.9066, 0.9340, and 0.9500. With this setting of location shifts, the required numbers of blocks are nearly identical. However, the attained power levels still vary with the underlying distributions. The practical usefulness of power analysis for generating important findings inevitably relies on a good understanding of research contexts and data structures. The proposed test procedure and power function are computationally simple and will give accurate outputs when all the necessary information is properly specified.

%4
\section{Conclusions}  
In view of the limitations of the existing methods, this paper identifies a general class of $F$-transformation of the Friedman statistic. Two prominent members are identified and compared with the current methods for the adequacy in Type I error control. Moreover, two approximate noncentral $F$ distributions are presented for the proposed $F$-transformation under the alternative hypothesis with certain location shifts. Explicit expressions are derived to facilitate the developed noncentral $F$ approximations for power calculations when the underlying populations have uniform, normal, Laplace, and exponential distributions. Monte Carlo simulation results demonstrated that the recommended power function is more accurate than the other contending methods including the previously documented method based on noncentral chi-square approximation. Consequently, the proposed $F$-transformation test and associated power function provide useful alternatives to the Friedman test and power procedure for detecting location shifts in randomized complete block design. 

\newpage
%\subsection*{Statements and Declarations}

\subsubsection*{Declaration of Conflicting Interests} 
\noindent 
The author(s) declared no potential conflicts of interest with respect to the research, authorship, and/or publication of this article. 

\subsubsection*{Data availability} 
\noindent 
The datasets generated during and/or analyzed during the current study are available in the cited references.

\subsubsection*{Funding} 
\noindent 
This study was funded by Ministry of Science and Technology.

\newpage
\subsection*{Appendix}

Exact probability evaluations of the uniform, normal, Laplace, and exponential distributions
\\
Assume the location shifts $(\uptheta_1, \, \uptheta_2, \, \ldots, \, \uptheta_K)$ are $\uptheta_1 \leq \uptheta_2 \, \ldots \, \leq \uptheta_K$.
\\
1. Uniform $(-1/2, \, 1/2)$: $max\lvert \uptheta_i-\uptheta_l \rvert < 1$ for all $i$ and $l$
\begin{align*}
P_{1il}&=
\begin{cases}
\displaystyle 1/2+(\uptheta_i-\uptheta_l)+(1/2)(\uptheta_i-\uptheta_l)^2 \quad \mathrm{if} \ i<l
\\
\displaystyle 1/2+(\uptheta_i-\uptheta_l)-(1/2)(\uptheta_i-\uptheta_l)^2 \quad \mathrm{if} \ i>l
\end{cases}
\\
P_{2i(l,\,m)}&=
\begin{cases}
\displaystyle (1/3)-(1/2)(\uptheta_l-\uptheta_m)+(\uptheta_i-\uptheta_m)-(\uptheta_i-\uptheta_m)(\uptheta_l-\uptheta_m) 
\\
\quad +(\uptheta_i-\uptheta_m)^2-(1/2)(\uptheta_l-\uptheta_m)(\uptheta_i-\uptheta_m)^2+(1/3)(\uptheta_i-\uptheta_m)^3 \quad \mathrm{if} \ i<l<m
\\
\displaystyle (1/3)+(\uptheta_i-\uptheta_m)-(1/2)(\uptheta_l-\uptheta_m)+(1/2)(\uptheta_i-\uptheta_m)^2
\\
\quad -(1/2)(\uptheta_l-\uptheta_m)^2-(1/6)(\uptheta_l-\uptheta_m)^3 \quad \mathrm{if} \ l<i<m
\\
\displaystyle (1/3)+(\uptheta_i-\uptheta_m)-(1/2)(\uptheta_l-\uptheta_m)-(1/2)(\uptheta_l-\uptheta_m)(\uptheta_i-\uptheta_l)^2
\\
\quad -(1/2)(\uptheta_m-\uptheta_l)^2+(1/3)(\uptheta_l-\uptheta_i)^3 \quad \mathrm{if} \ l<m<i
\end{cases}
\end{align*}
\\
2. Normal (0, 1)
\begin{align*}
P_{1il}&=\Phi\lbrace(\uptheta_i-\uptheta_l)/2^{1/2}\rbrace  
\\
P_{2i(l,\,m)}&=E[\Phi(Z+\uptheta_i-\uptheta_l)\Phi(Z+\uptheta_i-\uptheta_m)] \, \mathrm{where} \, Z\,\sim\,N(0, 1).
\end{align*}
\\
3. Laplace (0, 1)
\begin{align*}
P_{1il}&=
\begin{cases}
\displaystyle (1/4)(2-\uptheta_i+\uptheta_l)exp(\uptheta_i-\uptheta_l) \quad \mathrm{if} \ i<l
\\
\displaystyle 1-(1/4)(2+\uptheta_i-\uptheta_l)exp(\uptheta_l-\uptheta_i) \quad \mathrm{if} \ i>l
\end{cases}
\\
P_{2i(l,\,m)}&=
\begin{cases}
\displaystyle (1/8)\lbrace{3-2(\uptheta_l-\uptheta_m)}\rbrace exp(\uptheta_i-\uptheta_m)-(1/12)exp(2\uptheta_i-\uptheta_l-\uptheta_m)
\\
\quad + (1/24)exp(\uptheta_i + \uptheta_l - 2\uptheta_m) \quad \mathrm{if} \ i<l<m
\\
\displaystyle (1/4)(2-\uptheta_i+\uptheta_m)exp(\uptheta_i-\uptheta_m)-(1/4)exp(\uptheta_l-\uptheta_m)
\\
\quad +(1/24)exp(\uptheta_i+\uptheta_l-2\uptheta_m)+(1/24)exp(2\uptheta_l-\uptheta_i-\uptheta_m) \quad \mathrm{if} \ l<i<m
\\
\displaystyle 1-(1/8)\lbrace{1+2(\uptheta_i-\uptheta_m)}\rbrace exp(\uptheta_l-\uptheta_i)-(1/4)(2+\uptheta_i-\uptheta_m)exp(\uptheta_m-\uptheta_i)
\\
\quad -(1/12)exp(\uptheta_l+\uptheta_m-2\uptheta_i)+(1/24)exp(2\uptheta_l-\uptheta_i-\uptheta_m) \quad \mathrm{if} \ l<m<i
\end{cases}
\end{align*}
\\
4. Exponential (1)
\begin{align*}
P_{1il}&=
\begin{cases}
\displaystyle (1/2)exp(\uptheta_i-\uptheta_l) \quad \mathrm{if} \ i<l
\\
\displaystyle 1-(1/2)exp(\uptheta_l-\uptheta_i) \quad \mathrm{if} \ i>l
\end{cases}
\\
P_{2i(l,\,m)}&=
\begin{cases}
\displaystyle (1/2)exp(\uptheta_i-\uptheta_m)-(1/6)exp(\uptheta_i+\uptheta_l-2\uptheta_m) \quad \mathrm{if} \ i<l<m \ \mathrm{or} \ l<i<m
\\
\displaystyle 1-(1/2)exp(\uptheta_l-\uptheta_i)-(1/2)exp(\uptheta_m-\uptheta_i)+(1/3)exp(\uptheta_l+\uptheta_m-2\uptheta_i) \quad \mathrm{if} \ l<m<i
\end{cases}
\end{align*}

\newpage
\bibliography{FMT-SANB2M}

\begin{thebibliography}{}

\bibitem[Box and Andersen, 1955]{Box1955}
Box, G. E.~P. and Andersen, S.~L. (1955).
\newblock Permutation theory in the derivation of robust criteria and the study of departures from assumption.
\newblock {\em Journal of the Royal Statistical Society, Series B}, 17:1--34.

\bibitem[Brownie and Boos, 1994]{Brownie1994}
Brownie, C. and Boos, D.~D. (1994).
\newblock Type {I} error robustness of {ANOVA} and {ANOVA} on ranks when the number of treatments is large.
\newblock {\em Biometrics}, 50:542--549.

\bibitem[Cochran and Cox, 1957]{Cochran1957}
Cochran, W.~G. and Cox, G.~M. (1957).
\newblock {\em Experimental Designs}.
\newblock Wiley, New York, 2nd edition.

\bibitem[Conover and Iman, 1981]{Conover1981}
Conover, W.~J. and Iman, R.~L. (1981).
\newblock Rank transformations as a bridge between parametric and nonparametric statistics.
\newblock {\em The American Statistician}, 35:124--129.

\bibitem[Friedman, 1937]{Friedman1937}
Friedman, M. (1937).
\newblock The use of ranks to avoid the assumption of normality implicit in the analysis of variance.
\newblock {\em Journal of the American Statistical Association}, 32:675--701.

\bibitem[Hettmansperger, 1984]{Hettmansperger1984}
Hettmansperger, T.~P. (1984).
\newblock {\em Statistical Inference Based on Ranks}.
\newblock Wiley, New York, NY.

\bibitem[Hollander and Wolfe, 1999]{Hollander1999}
Hollander, M. and Wolfe, D.~A. (1999).
\newblock {\em Nonparametric Statistical Methods}.
\newblock Wiley, New York, NY, 2nd edition.

\bibitem[Hollander et~al., 2013]{Hollander2013}
Hollander, M., Wolfe, D.~A., and Chicken, E. (2013).
\newblock {\em Nonparametric Statistical Methods}.
\newblock Wiley, New York, NY, 3rd edition.

\bibitem[Iman and Davenport, 1980]{Iman1980}
Iman, R.~L. and Davenport, J.~M. (1980).
\newblock Approximations of the critical region of the {Friedman} statistic.
\newblock {\em Communications in Statistics-Theory and Methods}, 9:571--595.

\bibitem[Iman et~al., 1984]{Iman1984}
Iman, R.~L., Hora, S.~C., and Conover, W.~J. (1984).
\newblock Comparison of asymptotically distribution-free procedures for the analysis of complete blocks.
\newblock {\em Journal of the American Statistical Association}, 79:674--685.

\bibitem[Kendall and Babington~Smith, 1939]{Kendall1939}
Kendall, M.~G. and Babington~Smith, B. (1939).
\newblock The problem of $m$ rankings.
\newblock {\em The Annals of Mathematical Statistics}, 10:275--287.

\bibitem[Lehmann, 1998]{Lehmann1998}
Lehmann, E.~L. (1998).
\newblock {\em Nonparametrics: Statistical Methods Based on Ranks}.
\newblock Prentice Hall, Upper Saddle River, New Jersey.

\bibitem[López-Vázquez and Hochsztain, 2019]{LopezVazquez2019}
López-Vázquez, C. and Hochsztain, E. (2019).
\newblock Extended and updated tables for the {Friedman} rank test.
\newblock {\em Communications in Statistics-Theory and Methods}, 48:268--281.

\bibitem[Mahfoud and Randles, 2005]{Mahfoud2005}
Mahfoud, Z.~R. and Randles, R.~H. (2005).
\newblock Practical tests for randomized complete block designs.
\newblock {\em Journal of Multivariate Analysis}, 96:73--92.

\bibitem[Martin et~al., 1993]{Martin1993}
Martin, L., Leblanc, R., and Toan, N.~K. (1993).
\newblock Tables for the {Friedman} rank test.
\newblock {\em Canadian Journal of Statistics}, 21:39--43.

\bibitem[Noether, 1987]{Noether1987}
Noether, G.~E. (1987).
\newblock Sample size determination for some common nonparametric tests.
\newblock {\em Journal of the American Statistical Association}, 82:645--647.

\bibitem[Odeh, 1977]{Odeh1977}
Odeh, R.~E. (1977).
\newblock Extended tables of the distribution of {Friedman’s $S$-statistic} in the two-way layout.
\newblock {\em Communications in Statistics-Simulation and Computation}, 6:29--48.

\bibitem[Owen, 1962]{Owen1962}
Owen, D.~B. (1962).
\newblock {\em Handbook of Statistical Tables}.
\newblock Addison-Wesley, Reading, MA.

\bibitem[Pitman, 1938]{Pitman1938}
Pitman, E. J.~G. (1938).
\newblock Significance tests which may be applied to samples from any populations: {III}. {The} analysis of variance test.
\newblock {\em Biometrika}, 29:322--335.

\bibitem[Schneider et~al., 2016]{Schneider2016}
Schneider, G., Chicken, E., and Becvarik, R. (2016).
\newblock {NSM3 Package: Functions and Datasets to Accompany Hollander, Wolfe, and Chicken–Nonparametric Statistical Methods, Third Edition}.

\bibitem[Serfling, 1980]{Serfling1980}
Serfling, R.~J. (1980).
\newblock {\em Approximation Theorems of Mathematical Statistics}.
\newblock Wiley, New York, NY.

\bibitem[Shieh et~al., 2006]{Shieh2006}
Shieh, G., Jan, S.~L., and Randles, R.~H. (2006).
\newblock On power and sample size determinations for the {Wilcoxon-Mann-Whitney} test.
\newblock {\em Journal of Nonparametric Statistics}, 18:33--43.

\bibitem[Shieh et~al., 2007]{Shieh2007}
Shieh, G., Jan, S.~L., and Randles, R.~H. (2007).
\newblock Power and sample size determinations for the {Wilcoxon} signed-rank test.
\newblock {\em Journal of Statistical Computation and Simulation}, 77:717--724.

\bibitem[Welch, 1937]{Welch1937}
Welch, B.~L. (1937).
\newblock On the $z$-test in randomized blocks and {Latin} squares.
\newblock {\em Biometrika}, 29:21--52.

\end{thebibliography}

\newpage
\linespread{1.5} 

\begin{landscape} 
%Table 1
\begin{table}[]
\caption{The difference between approximate and exact Type I error rates for $K = 3$.}
\label{tab_1}
\begin{tabular}{>{\raggedleft\arraybackslash}p{0.5cm}>{\raggedleft\arraybackslash}p{1.5cm}>{\raggedleft\arraybackslash}p{1.2cm}>{\centering\arraybackslash}p{0.2cm}>{\raggedleft\arraybackslash}p{1.8cm}>{\raggedleft\arraybackslash}p{1.8cm}>{\raggedleft\arraybackslash}p{1.8cm}>{\raggedleft\arraybackslash}p{1.8cm}>{\centering\arraybackslash}p{0.2cm}>{\raggedleft\arraybackslash}p{1.5cm}>{\raggedleft\arraybackslash}p{1.5cm}>{\raggedleft\arraybackslash}p{1.5cm}>{\raggedleft\arraybackslash}p{1.5cm}}
\hline
\rule[-8pt]{0pt}{24pt} 
 & & & & \multicolumn{4}{c}{Error} & & \multicolumn{4}{c}{Percentage of error} \\
\cline{5-8}
\cline{10-13}
\rule[-8pt]{0pt}{24pt} 
$B$ & $\upalpha$ & $c_\upalpha$ & & $T$ & $F_R$ & $F_M$ & $F_L$ & & $T$ & $F_R$ & $F_M$ & $F_L$ \\
\hline
\rule[0pt]{0pt}{14pt}
3 & 0.1944 & 4.667 &  & $-0.09743$ & $-0.14502$ & $-0.10696$ & $-0.11349$ &  & $-50.12$ & $-74.60$  & $-55.02$ & $-58.38$ \\
3 & 0.0278 & 6.0   &  & $0.02199$  & $*$        & $*$        & $-0.00220$ &  & $79.09$  & $*$       & $*$      & $-7.91$  \\
  &        &       &  &            &            &            &            &  &          &           &          &          \\
4 & 0.1250 & 4.5   &  & $-0.01960$ & $-0.04126$ & $-0.01561$ & $-0.02737$ &  & $-15.68$ & $-33.01$  & $-12.49$ & $-21.90$ \\
4 & 0.0417 & 6.5   &  & $-0.00293$ & $-0.03511$ & $-0.02636$ & $-0.01806$ &  & $-7.02$  & $-84.19$  & $-63.21$ & $-43.32$ \\
4 & 0.0046 & 8.0   &  & $0.01372$  & $*$        & $*$        & $0.00160$  &  & $298.17$ & $*$       & $*$      & $34.70$  \\
  &        &       &  &            &            &            &            &  &          &           &          &          \\
5 & 0.0934 & 5.2   &  & $-0.01913$ & $-0.04032$ & $-0.02375$ & $-0.02801$ &  & $-20.48$ & $-43.16$  & $-25.43$ & $-29.99$ \\
5 & 0.0394 & 6.4   &  & $0.00136$  & $-0.02260$ & $-0.01243$ & $-0.00965$ &  & $3.46$   & $-57.37$  & $-31.55$ & $-24.49$ \\
5 & 0.0085 & 8.4   &  & $0.00650$  & $-0.00784$ & $-0.00656$ & $-0.00195$ &  & $76.42$  & $-92.29$  & $-77.22$ & $-22.99$ \\
  &        &       &  &            &            &            &            &  &          &           &          &          \\
6 & 0.1416 & 4.333 &  & $-0.02704$ & $-0.03515$ & $-0.02126$ & $-0.03009$ &  & $-19.10$ & $-24.83$  & $-15.01$ & $-21.25$ \\
6 & 0.0521 & 6.333 &  & $-0.00966$ & $-0.02862$ & $-0.01939$ & $-0.01856$ &  & $-19.11$ & $-54.93$  & $-37.23$ & $-35.63$ \\
6 & 0.0120 & 8.333 &  & $0.00350$  & $-0.00934$ & $-0.00684$ & $-0.00344$ &  & $29.20$  & $-77.80$  & $-57.04$ & $-28.68$ \\
  &        &       &  &            &            &            &            &  &          &           &          &          \\
7 & 0.1118 & 4.571 &  & $-0.01010$ & $-0.01849$ & $-0.00717$ & $-0.01357$ &  & $-9.03$  & $-16.54$  & $-6.41$  & $-12.14$ \\
7 & 0.0515 & 6.0   &  & $-0.00171$ & $-0.01668$ & $-0.00776$ & $-0.00858$ &  & $-3.33$  & $-32.40$  & $-15.06$ & $-16.66$ \\
7 & 0.0162 & 8.0   &  & $0.00212$  & $-0.01000$ & $-0.00654$ & $-0.00408$ &  & $13.06$  & $-61.75$  & $-40.35$ & $-25.17$ \\
  &        &       &  &            &            &            &            &  &          &           &          &          \\
8 & 0.0789 & 5.25  &  & $-0.00646$ & $-0.01710$ & $-0.00803$ & $-0.01124$ &  & $-8.19$  & $-21.67$  & $-10.18$ & $-14.24$ \\
8 & 0.0469 & 6.25  &  & $-0.00296$ & $-0.01570$ & $-0.00854$ & $-0.00892$ &  & $-6.32$  & $-33.47$  & $-18.21$ & $-19.01$ \\
8 & 0.0099 & 9.0   &  & $0.00121$  & $-0.00683$ & $-0.00503$ & $-0.00312$ &  & $12.21$  & $-69.01$  & $-50.83$ & $-31.49$ \\
\hline
\end{tabular}
\end{table}
\end{landscape}

\begin{landscape} 
%Table 2
\begin{table}[]
\caption{The difference between approximate and exact Type I error rates for $K = 4$.}
\label{tab_2}
\begin{tabular}{>{\raggedleft\arraybackslash}p{0.5cm}>{\raggedleft\arraybackslash}p{1.5cm}>{\raggedleft\arraybackslash}p{1.2cm}>{\centering\arraybackslash}p{0.2cm}>{\raggedleft\arraybackslash}p{1.8cm}>{\raggedleft\arraybackslash}p{1.8cm}>{\raggedleft\arraybackslash}p{1.8cm}>{\raggedleft\arraybackslash}p{1.8cm}>{\centering\arraybackslash}p{0.2cm}>{\raggedleft\arraybackslash}p{1.5cm}>{\raggedleft\arraybackslash}p{1.5cm}>{\raggedleft\arraybackslash}p{1.5cm}>{\raggedleft\arraybackslash}p{1.5cm}}
\hline
\rule[-8pt]{0pt}{24pt} 
 & & & & \multicolumn{4}{c}{Error} & & \multicolumn{4}{c}{Percentage of error} \\
\cline{5-8}
\cline{10-13}
\rule[-8pt]{0pt}{24pt} 
$B$ & $\upalpha$ & $c_\upalpha$ & & $T$ & $F_R$ & $F_M$ & $F_L$ & & $T$ & $F_R$ & $F_M$ & $F_L$ \\
\hline
\rule[0pt]{0pt}{14pt}
3 & 0.1476 & 5.8    &  & $-0.02584$ & $-0.06349$ & $-0.03758$ & $-0.04434$ &  & $-17.51$ & $-43.01$ & $-25.46$ & $-30.04$ \\
3 & 0.0538 & 7.0    &  & $0.01810$  & $-0.03189$ & $-0.01631$ & $-0.00938$ &  & $33.64$  & $-59.28$ & $-30.32$ & $-17.44$ \\
3 & 0.0174 & 8.2    &  & $0.02465$  & $-0.01592$ & $-0.01290$ & $-0.00195$ &  & $141.69$ & $-91.47$ & $-74.15$ & $-11.23$ \\
  &        &        &  &            &            &            &            &  &          &          &          &          \\
4 & 0.1053 & 6.0    &  & $0.00631$  & $-0.01759$ & $-0.00083$ & $-0.00622$ &  & $5.99$   & $-16.70$ & $-0.78$  & $-5.91$  \\
4 & 0.0517 & 7.5    &  & $0.00586$  & $-0.02564$ & $-0.01477$ & $-0.01213$ &  & $11.33$  & $-49.59$ & $-28.58$ & $-23.46$ \\
4 & 0.0115 & 9.3    &  & $0.01406$  & $-0.00866$ & $-0.00584$ & $-0.00099$ &  & $122.24$ & $-75.30$ & $-50.82$ & $-8.63$  \\
  &        &        &  &            &            &            &            &  &          &          &          &          \\
5 & 0.1066 & 6.12   &  & $-0.00068$ & $-0.01823$ & $-0.00594$ & $-0.01019$ &  & $-0.64$  & $-17.10$ & $-5.57$  & $-9.56$  \\
5 & 0.0548 & 7.32   &  & $0.00757$  & $-0.01529$ & $-0.00563$ & $-0.00545$ &  & $13.81$  & $-27.89$ & $-10.27$ & $-9.94$  \\
5 & 0.0120 & 9.72   &  & $0.00910$  & $-0.00734$ & $-0.00453$ & $-0.00151$ &  & $75.85$  & $-61.17$ & $-37.72$ & $-12.58$ \\
  &        &        &  &            &            &            &            &  &          &          &          &          \\
6 & 0.1081 & 6.2    &  & $-0.00582$ & $-0.01970$ & $-0.01003$ & $-0.01349$ &  & $-5.39$  & $-18.22$ & $-9.27$  & $-12.48$ \\
6 & 0.0558 & 7.4    &  & $0.00438$  & $-0.01353$ & $-0.00570$ & $-0.00592$ &  & $7.86$   & $-24.25$ & $-10.21$ & $-10.60$ \\
6 & 0.0103 & 10.0   &  & $0.00827$  & $-0.00453$ & $-0.00201$ & $0.00012$  &  & $80.25$  & $-44.00$ & $-19.47$ & $1.20$   \\
  &        &        &  &            &            &            &            &  &          &          &          &          \\
7 & 0.1000 & 6.257  &  & $-0.00025$ & $-0.01172$ & $-0.00375$ & $-0.00667$ &  & $-0.25$  & $-11.72$ & $-3.75$  & $-6.67$  \\
7 & 0.0520 & 7.629  &  & $0.00233$  & $-0.01248$ & $-0.00620$ & $-0.00627$ &  & $4.49$   & $-24.00$ & $-11.92$ & $-12.06$ \\
7 & 0.0100 & 10.371 &  & $0.00566$  & $-0.00434$ & $-0.00230$ & $-0.00069$ &  & $56.62$  & $-43.36$ & $-22.95$ & $-6.85$  \\
  &        &        &  &            &            &            &            &  &          &          &          &          \\
8 & 0.1000 & 6.3    &  & $-0.00211$ & $-0.01189$ & $-0.00513$ & $-0.00764$ &  & $-2.11$  & $-11.89$ & $-5.13$  & $-7.64$  \\
8 & 0.0510 & 7.5    &  & $0.00656$  & $-0.00594$ & $-0.00030$ & $-0.00071$ &  & $12.86$  & $-11.64$ & $-0.58$  & $-1.39$  \\
8 & 0.0110 & 10.35  &  & $0.00481$  & $-0.00400$ & $-0.00201$ & $-0.00070$ &  & $43.76$  & $-36.34$ & $-18.24$ & $-6.39$  \\
\hline
\end{tabular}
\end{table}
\end{landscape}

\begin{landscape} 
%Table 3
\begin{table}[]
\caption{The difference between approximate and exact Type I error rates for $K = 5$.}
\label{tab_3}
\begin{tabular}{>{\raggedleft\arraybackslash}p{0.5cm}>{\raggedleft\arraybackslash}p{1.5cm}>{\raggedleft\arraybackslash}p{1.2cm}>{\centering\arraybackslash}p{0.2cm}>{\raggedleft\arraybackslash}p{1.8cm}>{\raggedleft\arraybackslash}p{1.8cm}>{\raggedleft\arraybackslash}p{1.8cm}>{\raggedleft\arraybackslash}p{1.8cm}>{\centering\arraybackslash}p{0.2cm}>{\raggedleft\arraybackslash}p{1.5cm}>{\raggedleft\arraybackslash}p{1.5cm}>{\raggedleft\arraybackslash}p{1.5cm}>{\raggedleft\arraybackslash}p{1.5cm}}
\hline
\rule[-8pt]{0pt}{24pt} 
 & & & & \multicolumn{4}{c}{Error} & & \multicolumn{4}{c}{Percentage of error} \\
\cline{5-8}
\cline{10-13}
\rule[-8pt]{0pt}{24pt} 
$B$ & $\upalpha$ & $c_\upalpha$ & & $T$ & $F_R$ & $F_M$ & $F_L$ & & $T$ & $F_R$ & $F_M$ & $F_L$ \\
\hline
\rule[0pt]{0pt}{14pt}
3 & 0.1172 & 7.2    &  & $0.00849$ & $-0.03016$ & $-0.01138$ & $-0.01478$ &  & $7.24$   & $-25.73$ & $-9.71$  & $-12.61$ \\
3 & 0.0559 & 8.267  &  & $0.02638$ & $-0.02072$ & $-0.00712$ & $-0.00363$ &  & $47.20$  & $-37.06$ & $-12.73$ & $-6.49$  \\
3 & 0.0151 & 9.867  &  & $0.02763$ & $-0.01082$ & $-0.00688$ & $-0.00019$ &  & $183.01$ & $-71.63$ & $-45.55$ & $-1.24$  \\
  &        &        &  &           &            &            &            &  &          &          &          &          \\
4 & 0.1129 & 7.4    &  & $0.00330$ & $-0.02187$ & $-0.00931$ & $-0.01235$ &  & $2.92$   & $-19.37$ & $-8.24$  & $-10.94$ \\
4 & 0.0597 & 8.6    &  & $0.01221$ & $-0.01835$ & $-0.00861$ & $-0.00757$ &  & $20.46$  & $-30.73$ & $-14.42$ & $-12.68$ \\
4 & 0.0102 & 11.0   &  & $0.01636$ & $-0.00543$ & $-0.00269$ & $0.00054$  &  & $160.43$ & $-53.21$ & $-26.35$ & $5.32$   \\
  &        &        &  &           &            &            &            &  &          &          &          &          \\
5 & 0.1070 & 7.52   &  & $0.00383$ & $-0.01487$ & $-0.00548$ & $-0.00799$ &  & $3.58$   & $-13.90$ & $-5.13$  & $-7.47$  \\
5 & 0.0560 & 8.80   &  & $0.01030$ & $-0.01228$ & $-0.00482$ & $-0.00443$ &  & $18.39$  & $-21.94$ & $-8.60$  & $-7.92$  \\
5 & 0.0100 & 11.52  &  & $0.01130$ & $-0.00414$ & $-0.00182$ & $0.00026$  &  & $113.02$ & $-41.42$ & $-18.19$ & $2.60$   \\
  &        &        &  &           &            &            &            &  &          &          &          &          \\
6 & 0.1025 & 7.60   &  & $0.00488$ & $-0.01001$ & $-0.00253$ & $-0.00463$ &  & $4.76$   & $-9.76$  & $-2.47$  & $-4.52$  \\
6 & 0.0550 & 8.933  &  & $0.00780$ & $-0.01010$ & $-0.00407$ & $-0.00393$ &  & $14.17$  & $-18.36$ & $-7.40$  & $-7.14$  \\
6 & 0.0109 & 11.733 &  & $0.00855$ & $-0.00373$ & $-0.00162$ & $-0.00010$ &  & $78.45$  & $-34.25$ & $-14.90$ & $-0.91$  \\
  &        &        &  &           &            &            &            &  &          &          &          &          \\
7 & 0.1025 & 7.657  &  & $0.00248$ & $-0.00988$ & $-0.00368$ & $-0.00547$ &  & $2.42$   & $-9.64$  & $-3.59$  & $-5.34$  \\
7 & 0.0527 & 9.029  &  & $0.00768$ & $-0.00713$ & $-0.00208$ & $-0.00206$ &  & $14.57$  & $-13.53$ & $-3.96$  & $-3.90$  \\
7 & 0.0100 & 12.114 &  & $0.00652$ & $-0.00311$ & $-0.00141$ & $-0.00024$ &  & $65.23$  & $-31.12$ & $-14.12$ & $-2.38$  \\
  &        &        &  &           &            &            &            &  &          &          &          &          \\
8 & 0.1039 & 7.6    &  & $0.00348$ & $-0.00677$ & $-0.00144$ & $-0.00314$ &  & $3.35$   & $-6.53$  & $-1.39$  & $-3.02$  \\
8 & 0.0521 & 9.1    &  & $0.00655$ & $-0.00608$ & $-0.00175$ & $-0.00177$ &  & $12.57$  & $-11.67$ & $-3.35$  & $-3.40$  \\
8 & 0.0104 & 12.2   &  & $0.00552$ & $-0.00276$ & $-0.00120$ & $-0.00024$ &  & $53.12$  & $-26.53$ & $-11.58$ & $-2.31$  \\
\hline
\end{tabular}
\end{table}
\end{landscape}

\begin{landscape} 
%Table 4
\begin{table}[]
\caption{The difference between simulated and nominal Type I error rates for $K = 3$.}
\label{tab_4}
\begin{tabular}{>{\raggedleft\arraybackslash}p{0.5cm}>{\raggedleft\arraybackslash}p{1.5cm}>{\raggedleft\arraybackslash}p{1.2cm}>{\centering\arraybackslash}p{0.2cm}>{\raggedleft\arraybackslash}p{1.8cm}>{\raggedleft\arraybackslash}p{1.8cm}>{\raggedleft\arraybackslash}p{1.8cm}>{\raggedleft\arraybackslash}p{1.8cm}>{\centering\arraybackslash}p{0.2cm}>{\raggedleft\arraybackslash}p{1.5cm}>{\raggedleft\arraybackslash}p{1.5cm}>{\raggedleft\arraybackslash}p{1.5cm}>{\raggedleft\arraybackslash}p{1.5cm}}
\hline
\rule[-8pt]{0pt}{24pt} 
 & & & & \multicolumn{4}{c}{Error} & & \multicolumn{4}{c}{Percentage of error} \\
\cline{5-8}
\cline{10-13}
\rule[-8pt]{0pt}{24pt} 
$B$ & $\upalpha$ & & & $T$ & $F_R$ & $F_M$ & $F_L$ & & $T$ & $F_R$ & $F_M$ & $F_L$ \\
\hline
\rule[0pt]{0pt}{14pt}
5  & 0.10 &  &  & $0.02563$  & $0.02489$  & $0.02489$  & $0.02563$  &  & $25.63$  & $24.89$  & $24.89$  & $25.63$  \\
   & 0.05 &  &  & $-0.01045$ & $-0.01119$ & $-0.01119$ & $-0.01045$ &  & $-20.90$ & $-22.38$ & $-22.38$ & $-20.90$ \\
   & 0.01 &  &  & $-0.00926$ & $0.01350$  & $0.01350$  & $-0.00157$ &  & $-92.60$ & $135.00$ & $135.00$ & $-15.70$ \\
   &      &  &  &            &            &            &            &  &          &          &          &          \\
10 & 0.10 &  &  & $-0.00824$ & $-0.00824$ & $-0.00824$ & $-0.00824$ &  & $-8.24$  & $-8.24$  & $-8.24$  & $-8.24$  \\
   & 0.05 &  &  & $-0.00534$ & $-0.00534$ & $-0.00534$ & $-0.00534$ &  & $-10.68$ & $-10.68$ & $-10.68$ & $-10.68$ \\
   & 0.01 &  &  & $-0.00274$ & $0.00080$  & $0.00080$  & $0.00080$  &  & $-27.40$ & $8.00$   & $8.00$   & $8.00$   \\
   &      &  &  &            &            &            &            &  &          &          &          &          \\
15 & 0.10 &  &  & $0.00553$  & $0.00553$  & $0.00553$  & $0.00553$  &  & $5.53$   & $5.53$   & $5.53$   & $5.53$   \\
   & 0.05 &  &  & $-0.00357$ & $-0.00357$ & $-0.00357$ & $-0.00357$ &  & $-7.14$  & $-7.14$  & $-7.14$  & $-7.14$  \\
   & 0.01 &  &  & $-0.00286$ & $0.00097$  & $-0.00048$ & $-0.00048$ &  & $-28.60$ & $9.70$   & $-4.80$  & $-4.80$  \\
   &      &  &  &            &            &            &            &  &          &          &          &          \\
20 & 0.10 &  &  & $0.00309$  & $0.00309$  & $0.00309$  & $0.00309$  &  & $3.09$   & $3.09$   & $3.09$   & $3.09$   \\
   & 0.05 &  &  & $0.00146$  & $0.00146$  & $0.00146$  & $0.00146$  &  & $2.92$   & $2.92$   & $2.92$   & $2.92$   \\
   & 0.01 &  &  & $-0.00174$ & $0.00146$  & $0.00146$  & $0.00146$  &  & $-17.40$ & $14.60$  & $14.60$  & $14.60$  \\
\hline
\end{tabular}
\end{table}
\end{landscape}

\begin{landscape} 
%Table 5
\begin{table}[]
\caption{The difference between simulated and nominal Type I error rates for $K = 4$.}
\label{tab_5}
\begin{tabular}{>{\raggedleft\arraybackslash}p{0.5cm}>{\raggedleft\arraybackslash}p{1.5cm}>{\raggedleft\arraybackslash}p{1.2cm}>{\centering\arraybackslash}p{0.2cm}>{\raggedleft\arraybackslash}p{1.8cm}>{\raggedleft\arraybackslash}p{1.8cm}>{\raggedleft\arraybackslash}p{1.8cm}>{\raggedleft\arraybackslash}p{1.8cm}>{\centering\arraybackslash}p{0.2cm}>{\raggedleft\arraybackslash}p{1.5cm}>{\raggedleft\arraybackslash}p{1.5cm}>{\raggedleft\arraybackslash}p{1.5cm}>{\raggedleft\arraybackslash}p{1.5cm}}
\hline
\rule[-8pt]{0pt}{24pt} 
 & & & & \multicolumn{4}{c}{Error} & & \multicolumn{4}{c}{Percentage of error} \\
\cline{5-8}
\cline{10-13}
\rule[-8pt]{0pt}{24pt} 
$B$ & $\upalpha$ & & & $T$ & $F_R$ & $F_M$ & $F_L$ & & $T$ & $F_R$ & $F_M$ & $F_L$ \\
\hline
\rule[0pt]{0pt}{14pt}
5  & 0.10 &  &  & $-0.00496$ & $0.00845$  & $-0.00496$ & $0.00845$  &  & $-4.96$  & $8.45$   & $-4.96$ & $8.45$  \\
   & 0.05 &  &  & $-0.01558$ & $0.01756$  & $0.00570$  & $0.00570$  &  & $-31.16$ & $35.12$  & $11.40$ & $11.40$ \\
   & 0.01 &  &  & $-0.00826$ & $0.01022$  & $0.00271$  & $-0.00060$ &  & $-82.60$ & $102.20$ & $27.10$ & $-6.00$ \\
   &      &  &  &            &            &            &            &  &          &          &         &         \\
10 & 0.10 &  &  & $-0.00449$ & $0.00620$  & $0.00187$  & $0.00187$  &  & $-4.49$  & $6.20$   & $1.87$  & $1.87$  \\
   & 0.05 &  &  & $-0.00481$ & $0.00374$  & $-0.00242$ & $-0.00242$ &  & $-9.62$  & $7.48$   & $-4.84$ & $-4.84$ \\
   & 0.01 &  &  & $-0.00353$ & $0.00313$  & $0.00136$  & $-0.00019$ &  & $-35.30$ & $31.30$  & $13.60$ & $-1.90$ \\
   &      &  &  &            &            &            &            &  &          &          &         &         \\
15 & 0.10 &  &  & $-0.00510$ & $-0.00052$ & $-0.00510$ & $-0.00052$ &  & $-5.10$  & $-0.52$  & $-5.10$ & $-0.52$ \\
   & 0.05 &  &  & $-0.00456$ & $-0.00170$ & $-0.00170$ & $-0.00170$ &  & $-9.12$  & $-3.40$  & $-3.40$ & $-3.40$ \\
   & 0.01 &  &  & $-0.00259$ & $0.00132$  & $0.00065$  & $-0.00051$ &  & $-25.90$ & $13.20$  & $6.50$  & $-5.10$ \\
   &      &  &  &            &            &            &            &  &          &          &         &         \\
20 & 0.10 &  &  & $-0.00482$ & $-0.00412$ & $-0.00412$ & $-0.00412$ &  & $-4.82$  & $-4.12$  & $-4.12$ & $-4.12$ \\
   & 0.05 &  &  & $-0.00371$ & $-0.00089$ & $-0.00100$ & $-0.00100$ &  & $-7.42$  & $-1.78$  & $-2.00$ & $-2.00$ \\
   & 0.01 &  &  & $-0.00220$ & $0.00085$  & $0.00000$  & $-0.00009$ &  & $-22.00$ & $8.50$   & $0.00$  & $-0.90$ \\
\hline
\end{tabular}
\end{table}
\end{landscape}

\begin{landscape} 
%Table 6
\begin{table}[]
\caption{The difference between simulated and nominal Type I error rates for $K = 5$.}
\label{tab_6}
\begin{tabular}{>{\raggedleft\arraybackslash}p{0.5cm}>{\raggedleft\arraybackslash}p{1.5cm}>{\raggedleft\arraybackslash}p{1.2cm}>{\centering\arraybackslash}p{0.2cm}>{\raggedleft\arraybackslash}p{1.8cm}>{\raggedleft\arraybackslash}p{1.8cm}>{\raggedleft\arraybackslash}p{1.8cm}>{\raggedleft\arraybackslash}p{1.8cm}>{\centering\arraybackslash}p{0.2cm}>{\raggedleft\arraybackslash}p{1.5cm}>{\raggedleft\arraybackslash}p{1.5cm}>{\raggedleft\arraybackslash}p{1.5cm}>{\raggedleft\arraybackslash}p{1.5cm}}
\hline
\rule[-8pt]{0pt}{24pt} 
 & & & & \multicolumn{4}{c}{Error} & & \multicolumn{4}{c}{Percentage of error} \\
\cline{5-8}
\cline{10-13}
\rule[-8pt]{0pt}{24pt} 
$B$ & $\upalpha$ & & & $T$ & $F_R$ & $F_M$ & $F_L$ & & $T$ & $F_R$ & $F_M$ & $F_L$ \\
\hline
\rule[0pt]{0pt}{14pt}
5  & 0.10 &  &  & $-0.01148$ & $0.00813$  & $-0.00540$ & $0.00813$  &  & $-11.48$ & $8.13$  & $-5.40$ & $8.13$  \\
   & 0.05 &  &  & $-0.01615$ & $0.00699$  & $-0.00171$ & $-0.00171$ &  & $-32.30$ & $13.98$ & $-3.42$ & $-3.42$ \\
   & 0.01 &  &  & $-0.00704$ & $0.00373$  & $0.00212$  & $0.00004$  &  & $-70.40$ & $37.30$ & $21.20$ & $0.40$  \\
   &      &  &  &            &            &            &            &  &          &         &         &         \\
10 & 0.10 &  &  & $-0.00619$ & $0.00571$  & $-0.00078$ & $0.00067$  &  & $-6.19$  & $5.71$  & $-0.78$ & $0.67$  \\
   & 0.05 &  &  & $-0.00652$ & $0.00203$  & $0.00110$  & $0.00110$  &  & $-13.04$ & $4.06$  & $2.20$  & $2.20$  \\
   & 0.01 &  &  & $-0.00371$ & $0.00212$  & $0.00051$  & $-0.00046$ &  & $-37.10$ & $21.20$ & $5.10$  & $-4.60$ \\
   &      &  &  &            &            &            &            &  &          &         &         &         \\
15 & 0.10 &  &  & $-0.00444$ & $0.00012$  & $-0.00136$ & $-0.00136$ &  & $-4.44$  & $0.12$  & $-1.36$ & $-1.36$ \\
   & 0.05 &  &  & $-0.00531$ & $0.00063$  & $-0.00172$ & $-0.00172$ &  & $-10.62$ & $1.26$  & $-3.44$ & $-3.44$ \\
   & 0.01 &  &  & $-0.00238$ & $0.00096$  & $0.00001$  & $-0.00017$ &  & $-23.80$ & $9.60$  & $0.10$  & $-1.70$ \\
   &      &  &  &            &            &            &            &  &          &         &         &         \\
20 & 0.10 &  &  & $-0.00443$ & $-0.00028$ & $-0.00329$ & $-0.00028$ &  & $-4.43$  & $-0.28$ & $-3.29$ & $-0.28$ \\
   & 0.05 &  &  & $-0.00372$ & $0.00181$  & $-0.00061$ & $-0.00061$ &  & $-7.44$  & $3.62$  & $-1.22$ & $-1.22$ \\
   & 0.01 &  &  & $-0.00174$ & $0.00065$  & $-0.00005$ & $-0.00016$ &  & $-17.40$ & $6.50$  & $-0.50$ & $-1.60$ \\
\hline
\end{tabular}
\end{table}
\end{landscape}

\begin{landscape} 
%Table 7
\begin{table}[]
\caption{Estimated power and simulated power for uniform shift alternatives and $K = 3$.}
\label{tab_7}
%\vspace{8pt} 
\renewcommand{\arraystretch}{1} 
\begin{tabular}{>{\raggedright\arraybackslash}p{6.5cm}>{\raggedright\arraybackslash}p{1.5cm}>{\raggedleft\arraybackslash}p{1.5cm}>{\centering\arraybackslash}p{3cm}>{\centering\arraybackslash}p{3cm}>{\raggedleft\arraybackslash}p{2cm}}
\hline
\rule[-8pt]{0pt}{24pt} 
Shifts & Procedure & $B$ & Estimated power & Simulated power & Difference \\
\hline
\rule[0pt]{0pt}{14pt}
$0.2887 \cdot (-1, \, 0, \, 1)$ & $\hspace{12pt}T_H$        & 9  & 0.9186          & 0.8178          & $0.1007$     \\
                                & $\hspace{12pt}F_{MA}$     & 11 & 0.9146          & 0.8773          & $0.0373$     \\
                                & $\hspace{12pt}F_{MB}$     & 10 & 0.9214          & 0.8478          & $0.0736$     \\
                                & $\hspace{12pt}F_{LA}$     & 13 & 0.9069          & 0.9418          & $-0.0348$    \\
                                & $\hspace{12pt}F_{LB}$     & 13 & 0.9205          & 0.9411          & $-0.0207$    \\
                                &                           &    &                 &                 &              \\
$0.2887 \cdot (-2/3, \, 0, \, 2/3)$ & $\hspace{12pt}T_H$        & 19 & 0.9003          & 0.8259          & $0.0744$     \\
                                    & $\hspace{12pt}F_{MA}$     & 23 & 0.9023          & 0.8836          & $0.0187$     \\
                                    & $\hspace{12pt}F_{MB}$     & 22 & 0.9050          & 0.8673          & $0.0377$     \\
                                    & $\hspace{12pt}F_{LA}$     & 26 & 0.9119          & 0.9238          & $-0.0119$    \\
                                    & $\hspace{12pt}F_{LB}$     & 25 & 0.9070          & 0.9153          & $-0.0083$    \\
                                    &                           &    &                 &                 &              \\
$0.2887 \cdot (-1/2, \, 0, \, 1/2)$ & $\hspace{12pt}T_H$        & 34 & 0.9023          & 0.8367          & $0.0656$     \\
                                    & $\hspace{12pt}F_{MA}$     & 40 & 0.9039          & 0.8963          & $0.0076$     \\
                                    & $\hspace{12pt}F_{MB}$     & 39 & 0.9053          & 0.8882          & $0.0170$     \\
                                    & $\hspace{12pt}F_{LA}$     & 42 & 0.9023          & 0.9115          & $-0.0092$    \\
                                    & $\hspace{12pt}F_{LB}$     & 42 & 0.9065          & 0.9117          & $-0.0053$    \\
                                    &                           &    &                 &                 &              \\
$0.2887 \cdot (-1/3, \, 0, \, 1/3)$ & $\hspace{12pt}T_H$        & 76 & 0.9003          & 0.8556          & $0.0447$     \\
                                    & $\hspace{12pt}F_{MA}$     & 86 & 0.9013          & 0.8987          & $0.0026$     \\
                                    & $\hspace{12pt}F_{MB}$     & 85 & 0.9019          & 0.8932          & $0.0087$     \\
                                    & $\hspace{12pt}F_{LA}$     & 88 & 0.9006          & 0.9038          & $-0.0032$    \\
                                    & $\hspace{12pt}F_{LB}$     & 88 & 0.9026          & 0.9033          & $-0.0006$    \\
\hline
\end{tabular}
\end{table}
\end{landscape}

\begin{landscape} 
%Table 8
\begin{table}[]
\caption{Estimated power and simulated power for uniform shift alternatives and $K = 5$.}
\label{tab_8}
%\vspace{8pt} 
\renewcommand{\arraystretch}{1} 
\begin{tabular}{>{\raggedright\arraybackslash}p{6.5cm}>{\raggedright\arraybackslash}p{1.5cm}>{\raggedleft\arraybackslash}p{1.5cm}>{\centering\arraybackslash}p{3cm}>{\centering\arraybackslash}p{3cm}>{\raggedleft\arraybackslash}p{2cm}}
\hline
\rule[-8pt]{0pt}{24pt} 
Shifts & Procedure & $B$ & Estimated power & Simulated power & Difference \\
\hline
\rule[0pt]{0pt}{14pt}
$0.2887 \cdot (-1, \, -1/2, \, 0, \, 1/2, \, 1)$ & $\hspace{12pt}T_H$          & 8  & 0.9237          & 0.7903          & $0.1333$     \\
                                                 & $\hspace{12pt}F_{MA}$       & 10 & 0.9155          & 0.9109          & $0.0046$     \\
                                                 & $\hspace{12pt}F_{MB}$       & 9  & 0.9024          & 0.8623          & $0.0402$     \\
                                                 & $\hspace{12pt}F_{LA}$       & 11 & 0.9178          & 0.9410          & $-0.0232$    \\
                                                 & $\hspace{12pt}F_{LB}$       & 11 & 0.9292          & 0.9390          & $-0.0098$    \\
                                                 &                &    &                 &                 &              \\
$0.2887 \cdot (-2/3, \, -1/3, \, 0, \, 1/3, \, 2/3)$ & $\hspace{12pt}T_H$        & 17 & 0.9068          & 0.8190          & $0.0878$     \\
                                                     & $\hspace{12pt}F_{MA}$     & 21 & 0.9127          & 0.9079          & $0.0049$     \\
                                                     & $\hspace{12pt}F_{MB}$     & 20 & 0.9070          & 0.8916          & $0.0153$     \\
                                                     & $\hspace{12pt}F_{LA}$     & 22 & 0.9139          & 0.9224          & $-0.0085$    \\
                                                     & $\hspace{12pt}F_{LB}$     & 21 & 0.9053          & 0.9101          & $-0.0048$    \\
                                                     &              &    &                 &                 &              \\
$0.2887 \cdot (-1/2, \, -1/4, \, 0, \, 1/4, \, 1/2)$ & $\hspace{12pt}T_H$        & 30 & 0.9045          & 0.8379          & $0.0666$     \\
                                                     & $\hspace{12pt}F_{MA}$     & 35 & 0.9034          & 0.9005          & $0.0029$     \\
                                                     & $\hspace{12pt}F_{MB}$     & 35 & 0.9092          & 0.9012          & $0.0080$     \\
                                                     & $\hspace{12pt}F_{LA}$     & 36 & 0.9045          & 0.9105          & $-0.0060$    \\
                                                     & $\hspace{12pt}F_{LB}$     & 36 & 0.9082          & 0.9103          & $-0.0021$    \\
                                                     &              &    &                 &                 &              \\
$0.2887 \cdot (-1/3, \, -1/6, \, 0, \, 1/6, \, 1/3)$ & $\hspace{12pt}T_H$        & 67 & 0.9022          & 0.8593          & $0.0428$     \\
                                                     & $\hspace{12pt}F_{MA}$     & 75 & 0.9014          & 0.8981          & $0.0033$     \\
                                                     & $\hspace{12pt}F_{MB}$     & 75 & 0.9042          & 0.8998          & $0.0044$     \\
                                                     & $\hspace{12pt}F_{LA}$     & 76 & 0.9020          & 0.9055          & $-0.0035$    \\
                                                     & $\hspace{12pt}F_{LB}$     & 76 & 0.9038          & 0.9035          & $0.0003$     \\
\hline
\end{tabular}
\end{table}
\end{landscape}

\begin{landscape} 
%Table 9
\begin{table}[]
\caption{Estimated power and simulated power for normal shift alternatives and $K = 3$.}
\label{tab_9}
%\vspace{8pt} 
\renewcommand{\arraystretch}{1} 
\begin{tabular}{>{\raggedright\arraybackslash}p{6.5cm}>{\raggedright\arraybackslash}p{1.5cm}>{\raggedleft\arraybackslash}p{1.5cm}>{\centering\arraybackslash}p{3cm}>{\centering\arraybackslash}p{3cm}>{\raggedleft\arraybackslash}p{2cm}}
\hline
\rule[-8pt]{0pt}{24pt} 
Shifts & Procedure & $B$ & Estimated power & Simulated power & Difference \\
\hline
\rule[0pt]{0pt}{14pt}
$(-1, \, 0, \, 1)$ & $\hspace{12pt}T_H$          & 9  & 0.9056          & 0.8505          & $0.0550$     \\
                   & $\hspace{12pt}F_{MA}$       & 10 & 0.9165          & 0.8803          & $0.0362$     \\
                   & $\hspace{12pt}F_{MB}$       & 9  & 0.9243          & 0.8519          & $0.0724$     \\
                   & $\hspace{12pt}F_{LA}$       & 12 & 0.9077          & 0.9494          & $-0.0416$    \\
                   & $\hspace{12pt}F_{LB}$       & 12 & 0.9224          & 0.9503          & $-0.0280$    \\
                   &                             &    &                 &                 &              \\
$(-2/3, \, 0, \, 2/3)$ & $\hspace{12pt}T_H$          & 20 & 0.9019          & 0.8786          & $0.0232$     \\
                       & $\hspace{12pt}F_{MA}$       & 21 & 0.9096          & 0.8996          & $0.0100$     \\
                       & $\hspace{12pt}F_{MB}$       & 20 & 0.9125          & 0.8795          & $0.0329$     \\
                       & $\hspace{12pt}F_{LA}$       & 23 & 0.9060          & 0.9163          & $-0.0104$    \\
                       & $\hspace{12pt}F_{LB}$       & 22 & 0.9002          & 0.9048          & $-0.0046$    \\
                       &                             &    &                 &                 &              \\
$(-1/2, \, 0, \, 1/2)$ & $\hspace{12pt}T_H$          & 36 & 0.9056          & 0.8980          & $0.0075$     \\
                       & $\hspace{12pt}F_{MA}$       & 36 & 0.9021          & 0.8974          & $0.0047$     \\
                       & $\hspace{12pt}F_{MB}$       & 35 & 0.9036          & 0.8801          & $0.0236$     \\
                       & $\hspace{12pt}F_{LA}$       & 38 & 0.9004          & 0.9039          & $-0.0034$    \\
                       & $\hspace{12pt}F_{LB}$       & 38 & 0.9051          & 0.9045          & $0.0006$     \\
                       &                             &    &                 &                 &              \\
$(-1/3, \, 0, \, 1/3)$ & $\hspace{12pt}T_H$          & 80 & 0.9019          & 0.8962          & $0.0056$     \\
                       & $\hspace{12pt}F_{MA}$       & 80 & 0.9005          & 0.8992          & $0.0013$     \\
                       & $\hspace{12pt}F_{MB}$       & 79 & 0.9011          & 0.8953          & $0.0058$     \\
                       & $\hspace{12pt}F_{LA}$       & 83 & 0.9035          & 0.9019          & $0.0016$     \\
                       & $\hspace{12pt}F_{LB}$       & 82 & 0.9019          & 0.9048          & $-0.0029$    \\
\hline
\end{tabular}
\end{table}
\end{landscape}

\begin{landscape} 
%Table 10
\begin{table}[]
\caption{Estimated power and simulated power for normal shift alternatives and $K = 5$.}
\label{tab_10}
%\vspace{8pt} 
\renewcommand{\arraystretch}{1} 
\begin{tabular}{>{\raggedright\arraybackslash}p{6.5cm}>{\raggedright\arraybackslash}p{1.5cm}>{\raggedleft\arraybackslash}p{1.5cm}>{\centering\arraybackslash}p{3cm}>{\centering\arraybackslash}p{3cm}>{\raggedleft\arraybackslash}p{2cm}}
\hline
\rule[-8pt]{0pt}{24pt} 
Shifts & Procedure & $B$ & Estimated power & Simulated power & Difference \\
\hline
\rule[0pt]{0pt}{14pt}
$(-1, \, -1/2, \, 0, \, 1/2, \, 1)$ & $\hspace{12pt}T_H$          & 8  & 0.9102          & 0.8368          & $0.0735$     \\
                                    & $\hspace{12pt}F_{MA}$       & 9  & 0.9145          & 0.8965          & $0.0180$     \\
                                    & $\hspace{12pt}F_{MB}$       & 9  & 0.9365          & 0.8958          & $0.0408$     \\
                                    & $\hspace{12pt}F_{LA}$       & 10 & 0.9173          & 0.9373          & $-0.0200$    \\
                                    & $\hspace{12pt}F_{LB}$       & 10 & 0.9300          & 0.9365          & $-0.0065$    \\
                                    &                             &    &                 &                 &              \\
$(-2/3, \, -1/3, \, 0, \, 1/3, \, 2/3)$ & $\hspace{12pt}T_H$          & 18 & 0.9102          & 0.8812          & $0.0291$     \\
                                        & $\hspace{12pt}F_{MA}$       & 19 & 0.9139          & 0.9068          & $0.0071$     \\
                                        & $\hspace{12pt}F_{MB}$       & 18 & 0.9075          & 0.8880          & $0.0195$     \\
                                        & $\hspace{12pt}F_{LA}$       & 20 & 0.9151          & 0.9230          & $-0.0079$    \\
                                        & $\hspace{12pt}F_{LB}$       & 19 & 0.9056          & 0.9082          & $-0.0026$    \\
                                        &                             &    &                 &                 &              \\
$(-1/2, \, -1/4, \, 0, \, 1/4, \, 1/2)$ & $\hspace{12pt}T_H$          & 31 & 0.9003          & 0.8825          & $0.0178$     \\
                                        & $\hspace{12pt}F_{MA}$       & 32 & 0.9027          & 0.8982          & $0.0045$     \\
                                        & $\hspace{12pt}F_{MB}$       & 32 & 0.9091          & 0.8990          & $0.0101$     \\
                                        & $\hspace{12pt}F_{LA}$       & 33 & 0.9039          & 0.9091          & $-0.0052$    \\
                                        & $\hspace{12pt}F_{LB}$       & 33 & 0.9080          & 0.9070          & $0.0010$     \\
                                        &                             &    &                 &                 &              \\
$(-1/3, \, -1/6, \, 0, \, 1/6, \, 1/3)$ & $\hspace{12pt}T_H$          & 70 & 0.9014          & 0.8948          & $0.0066$     \\
                                        & $\hspace{12pt}F_{MA}$       & 71 & 0.9026          & 0.8993          & $0.0033$     \\
                                        & $\hspace{12pt}F_{MB}$       & 70 & 0.9009          & 0.8968          & $0.0041$     \\
                                        & $\hspace{12pt}F_{LA}$       & 72 & 0.9032          & 0.9051          & $-0.0020$    \\
                                        & $\hspace{12pt}F_{LB}$       & 71 & 0.9005          & 0.9004          & $0.0001$     \\
\hline
\end{tabular}
\end{table}
\end{landscape}

\begin{landscape} 
%Table 11
\begin{table}[]
\caption{Estimated power and simulated power for Laplace shift alternatives and $K = 3$.}
\label{tab_11}
%\vspace{8pt} 
\renewcommand{\arraystretch}{1} 
\begin{tabular}{>{\raggedright\arraybackslash}p{6.5cm}>{\raggedright\arraybackslash}p{1.5cm}>{\raggedleft\arraybackslash}p{1.5cm}>{\centering\arraybackslash}p{3cm}>{\centering\arraybackslash}p{3cm}>{\raggedleft\arraybackslash}p{2cm}}
\hline
\rule[-8pt]{0pt}{24pt} 
Shifts & Procedure & $B$ & Estimated power & Simulated power & Difference \\
\hline
\rule[0pt]{0pt}{14pt}
$1.4142 \cdot (-1, \, 0, \, 1)$ & $\hspace{12pt}T_H$          & 6  & 0.9186          & 0.6989          & $0.2196$     \\
                                & $\hspace{12pt}F_{MA}$       & 9  & 0.9369          & 0.8963          & $0.0406$     \\
                                & $\hspace{12pt}F_{MB}$       & 7  & 0.9076          & 0.7898          & $0.1177$     \\
                                & $\hspace{12pt}F_{LA}$       & 11 & 0.9220          & 0.9398          & $-0.0178$    \\
                                & $\hspace{12pt}F_{LB}$       & 10 & 0.9101          & 0.9190          & $-0.0088$    \\
                                &                             &    &                 &                 &              \\
$1.4142 \cdot (-2/3, \, 0, \, 2/3)$ & $\hspace{12pt}T_H$          & 13 & 0.9080          & 0.7956          & $0.1124$     \\
                                    & $\hspace{12pt}F_{MA}$       & 16 & 0.9071          & 0.8910          & $0.0161$     \\
                                    & $\hspace{12pt}F_{MB}$       & 15 & 0.9113          & 0.8489          & $0.0624$     \\
                                    & $\hspace{12pt}F_{LA}$       & 18 & 0.9028          & 0.9117          & $-0.0090$    \\
                                    & $\hspace{12pt}F_{LB}$       & 18 & 0.9127          & 0.9104          & $0.0023$     \\
                                    &                             &    &                 &                 &              \\
$1.4142 \cdot (-1/2, \, 0, \, 1/2)$ & $\hspace{12pt}T_H$          & 23 & 0.9066          & 0.8294          & $0.0773$     \\
                                    & $\hspace{12pt}F_{MA}$       & 26 & 0.9003          & 0.8805          & $0.0198$     \\
                                    & $\hspace{12pt}F_{MB}$       & 25 & 0.9027          & 0.8695          & $0.0332$     \\
                                    & $\hspace{12pt}F_{LA}$       & 29 & 0.9092          & 0.9172          & $-0.0080$    \\
                                    & $\hspace{12pt}F_{LB}$       & 28 & 0.9048          & 0.9078          & $-0.0030$    \\
                                    &                             &    &                 &                 &              \\
$1.4142 \cdot (-1/3, \, 0, \, 1/3)$ & $\hspace{12pt}T_H$          & 51 & 0.9023          & 0.8642          & $0.0381$     \\
                                    & $\hspace{12pt}F_{MA}$       & 55 & 0.9012          & 0.8923          & $0.0090$     \\
                                    & $\hspace{12pt}F_{MB}$       & 54 & 0.9022          & 0.8881          & $0.0141$     \\
                                    & $\hspace{12pt}F_{LA}$       & 57 & 0.9002          & 0.9058          & $-0.0056$    \\
                                    & $\hspace{12pt}F_{LB}$       & 57 & 0.9033          & 0.9057          & $-0.0024$    \\
\hline
\end{tabular}
\end{table}
\end{landscape}

\begin{landscape} 
%Table 12
\begin{table}[]
\caption{Estimated power and simulated power for Laplace shift alternatives and $K = 5$.}
\label{tab_12}
%\vspace{8pt} 
\renewcommand{\arraystretch}{1} 
\begin{tabular}{>{\raggedright\arraybackslash}p{6.5cm}>{\raggedright\arraybackslash}p{1.5cm}>{\raggedleft\arraybackslash}p{1.5cm}>{\centering\arraybackslash}p{3cm}>{\centering\arraybackslash}p{3cm}>{\raggedleft\arraybackslash}p{2cm}}
\hline
\rule[-8pt]{0pt}{24pt} 
Shifts & Procedure & $B$ & Estimated power & Simulated power & Difference \\
\hline
\rule[0pt]{0pt}{14pt}
$1.4142 \cdot (-1, \, -1/2, \, 0, \, 1/2, \, 1)$ & $\hspace{12pt}T_H$          & 5  & 0.9045          & 0.6231          & $0.2814$     \\
                                                 & $\hspace{12pt}F_{MA}$       & 8  & 0.9399          & 0.9132          & $0.0267$     \\
                                                 & $\hspace{12pt}F_{MB}$       & 7  & 0.9253          & 0.8590          & $0.0663$     \\
                                                 & $\hspace{12pt}F_{LA}$       & 9  & 0.9386          & 0.9444          & $-0.0058$    \\
                                                 & $\hspace{12pt}F_{LB}$       & 8  & 0.9182          & 0.9121          & $0.0061$     \\
                                                 &                             &    &                 &                 &              \\
$1.4142 \cdot (-2/3, \, -1/3, \, 0, \, 1/3, \, 2/3)$ & $\hspace{12pt}T_H$          & 12 & 0.9237          & 0.8194          & $0.1042$     \\
                                                     & $\hspace{12pt}F_{MA}$       & 14 & 0.9071          & 0.8916          & $0.0155$     \\
                                                     & $\hspace{12pt}F_{MB}$       & 14 & 0.9218          & 0.8944          & $0.0274$     \\
                                                     & $\hspace{12pt}F_{LA}$       & 15 & 0.9096          & 0.9130          & $-0.0034$    \\
                                                     & $\hspace{12pt}F_{LB}$       & 15 & 0.9184          & 0.9133          & $0.0051$     \\
                                                     &                             &    &                 &                 &              \\
$1.4142 \cdot (-1/2, \, -1/4, \, 0, \, 1/4, \, 1/2)$ & $\hspace{12pt}T_H$          & 20 & 0.9045          & 0.8330          & $0.0716$     \\
                                                     & $\hspace{12pt}F_{MA}$       & 23 & 0.9047          & 0.8942          & $0.0105$     \\
                                                     & $\hspace{12pt}F_{MB}$       & 23 & 0.9136          & 0.8930          & $0.0206$     \\
                                                     & $\hspace{12pt}F_{LA}$       & 24 & 0.9063          & 0.9082          & $-0.0019$    \\
                                                     & $\hspace{12pt}F_{LB}$       & 24 & 0.9118          & 0.9086          & $0.0032$     \\
                                                     &                             &    &                 &                 &              \\
$1.4142 \cdot (-1/3, \, -1/6, \, 0, \, 1/6, \, 1/3)$ & $\hspace{12pt}T_H$          & 45 & 0.9045          & 0.8710          & $0.0336$     \\
                                                     & $\hspace{12pt}F_{MA}$       & 48 & 0.9012          & 0.8963          & $0.0049$     \\
                                                     & $\hspace{12pt}F_{MB}$       & 48 & 0.9055          & 0.8944          & $0.0111$     \\
                                                     & $\hspace{12pt}F_{LA}$       & 49 & 0.9021          & 0.9030          & $-0.0009$    \\
                                                     & $\hspace{12pt}F_{LB}$       & 49 & 0.9048          & 0.9040          & $0.0008$     \\
\hline
\end{tabular}
\end{table}
\end{landscape}

\begin{landscape} 
%Table 13
\begin{table}[]
\caption{Estimated power and simulated power for exponential shift alternatives and $K = 3$.}
\label{tab_13}
%\vspace{8pt} 
\renewcommand{\arraystretch}{1} 
\begin{tabular}{>{\raggedright\arraybackslash}p{6.5cm}>{\raggedright\arraybackslash}p{1.5cm}>{\raggedleft\arraybackslash}p{1.5cm}>{\centering\arraybackslash}p{3cm}>{\centering\arraybackslash}p{3cm}>{\raggedleft\arraybackslash}p{2cm}}
\hline
\rule[-8pt]{0pt}{24pt} 
Shifts & Procedure & $B$ & Estimated power & Simulated power & Difference \\
\hline
\rule[0pt]{0pt}{14pt}
$(-1, \, 0, \, 1)$ & $\hspace{12pt}T_H$          & 3  & 0.9186          & 0.2670          & $0.6516$     \\
                   & $\hspace{12pt}F_{MA}$       & 8  & 0.9366          & 0.8779          & $0.0587$     \\
                   & $\hspace{12pt}F_{MB}$       & 6  & 0.9029          & 0.7327          & $0.1702$     \\
                   & $\hspace{12pt}F_{LA}$       & 10 & 0.9201          & 0.9450          & $-0.0249$    \\
                   & $\hspace{12pt}F_{LB}$       & 9  & 0.9067          & 0.9242          & $-0.0175$    \\
                   &                             &    &                 &                 &              \\
$(-2/3, \, 0, \, 2/3)$ & $\hspace{12pt}T_H$          & 7  & 0.9280          & 0.5739          & $0.3542$     \\
                       & $\hspace{12pt}F_{MA}$       & 13 & 0.9041          & 0.8699          & $0.0342$     \\
                       & $\hspace{12pt}F_{MB}$       & 12 & 0.9099          & 0.8540          & $0.0558$     \\
                       & $\hspace{12pt}F_{LA}$       & 16 & 0.9196          & 0.9427          & $-0.0231$    \\
                       & $\hspace{12pt}F_{LB}$       & 15 & 0.9118          & 0.9113          & $0.0005$     \\
                       &                             &    &                 &                 &              \\
$(-1/2, \, 0, \, 1/2)$ & $\hspace{12pt}T_H$          & 12 & 0.9186          & 0.6702          & $0.2483$     \\
                       & $\hspace{12pt}F_{MA}$       & 20 & 0.9003          & 0.8803          & $0.0201$     \\
                       & $\hspace{12pt}F_{MB}$       & 19 & 0.9036          & 0.8672          & $0.0364$     \\
                       & $\hspace{12pt}F_{LA}$       & 23 & 0.9116          & 0.9179          & $-0.0063$    \\
                       & $\hspace{12pt}F_{LB}$       & 22 & 0.9060          & 0.9061          & $-0.0001$    \\
                       &                             &    &                 &                 &              \\
$(-1/3, \, 0, \, 1/3)$ & $\hspace{12pt}T_H$          & 26 & 0.9080          & 0.7201          & $0.1879$     \\
                       & $\hspace{12pt}F_{MA}$       & 39 & 0.9034          & 0.8938          & $0.0095$     \\
                       & $\hspace{12pt}F_{MB}$       & 38 & 0.9048          & 0.8782          & $0.0266$     \\
                       & $\hspace{12pt}F_{LA}$       & 41 & 0.9018          & 0.9084          & $-0.0067$    \\
                       & $\hspace{12pt}F_{LB}$       & 41 & 0.9061          & 0.9097          & $-0.0037$    \\
\hline
\end{tabular}
\end{table}
\end{landscape}

\begin{landscape} 
%Table 13
\begin{table}[]
\caption{Estimated power and simulated power for exponential shift alternatives and $K = 5$.}
\label{tab_13}
%\vspace{8pt} 
\renewcommand{\arraystretch}{1} 
\begin{tabular}{>{\raggedright\arraybackslash}p{6.5cm}>{\raggedright\arraybackslash}p{1.5cm}>{\raggedleft\arraybackslash}p{1.5cm}>{\centering\arraybackslash}p{3cm}>{\centering\arraybackslash}p{3cm}>{\raggedleft\arraybackslash}p{2cm}}
\hline
\rule[-8pt]{0pt}{24pt} 
Shifts & Procedure & $B$ & Estimated power & Simulated power & Difference \\
\hline
\rule[0pt]{0pt}{14pt}
$(-1, \, -1/2, \, 0, \, 1/2, \, 1)$ & $\hspace{12pt}T_H$          & 3  & 0.9520          & 0.2604          & $0.6916$     \\
                                    & $\hspace{12pt}F_{MA}$       & 7  & 0.9438          & 0.9076          & $0.0362$     \\
                                    & $\hspace{12pt}F_{MB}$       & 6  & 0.9269          & 0.8424          & $0.0845$     \\
                                    & $\hspace{12pt}F_{LA}$       & 8  & 0.9416          & 0.9478          & $-0.0063$    \\
                                    & $\hspace{12pt}F_{LB}$       & 7  & 0.9185          & 0.9074          & $0.0111$     \\
                                    &                             &    &                 &                 &              \\
$(-2/3, \, -1/3, \, 0, \, 1/3, \, 2/3)$ & $\hspace{12pt}T_H$          & 6  & 0.9237          & 0.5158          & $0.4078$     \\
                                        & $\hspace{12pt}F_{MA}$       & 11 & 0.9022          & 0.8830          & $0.0192$     \\
                                        & $\hspace{12pt}F_{MB}$       & 11 & 0.9217          & 0.8809          & $0.0407$     \\
                                        & $\hspace{12pt}F_{LA}$       & 12 & 0.9060          & 0.9125          & $-0.0065$    \\
                                        & $\hspace{12pt}F_{LB}$       & 12 & 0.9174          & 0.9115          & $0.0059$     \\
                                        &                             &    &                 &                 &              \\
$(-1/2, \, -1/4, \, 0, \, 1/4, \, 1/2)$ & $\hspace{12pt}T_H$          & 10 & 0.9045          & 0.6023          & $0.3022$     \\
                                        & $\hspace{12pt}F_{MA}$       & 17 & 0.9005          & 0.8849          & $0.0156$     \\
                                        & $\hspace{12pt}F_{MB}$       & 17 & 0.9130          & 0.8856          & $0.0274$     \\
                                        & $\hspace{12pt}F_{LA}$       & 18 & 0.9031          & 0.9078          & $-0.0047$    \\
                                        & $\hspace{12pt}F_{LB}$       & 18 & 0.9107          & 0.9067          & $0.0040$     \\
                                        &                             &    &                 &                 &              \\
$(-1/3, \, -1/6, \, 0, \, 1/6, \, 1/3)$ & $\hspace{12pt}T_H$          & 23 & 0.9113          & 0.7310          & $0.1804$     \\
                                        & $\hspace{12pt}F_{MA}$       & 33 & 0.9018          & 0.8938          & $0.0080$     \\
                                        & $\hspace{12pt}F_{MB}$       & 33 & 0.9081          & 0.8953          & $0.0128$     \\
                                        & $\hspace{12pt}F_{LA}$       & 34 & 0.9031          & 0.9057          & $-0.0027$    \\
                                        & $\hspace{12pt}F_{LB}$       & 34 & 0.9070          & 0.9058          & $0.0012$     \\
\hline
\end{tabular}
\end{table}
\end{landscape}

\end{document}